\pdfoutput=1
\documentclass[a4paper,12pt]{article}
\usepackage{amsfonts}
\usepackage{mathrsfs}
\usepackage{amsmath}
\usepackage{amssymb}
\usepackage[medium]{titlesec}
\usepackage{bm}
\usepackage{cite}
\usepackage[normalem]{ulem}
\usepackage{extarrows}
\usepackage{slashed}
\usepackage{isodateo}
\usepackage{graphicx}
\usepackage{xcolor}
\usepackage[bookmarksnumbered=true,bookmarksopen=true]{hyperref}
 \hypersetup{colorlinks,%
             linkcolor=[rgb]{0,0.2,0.6}, %
             citecolor=[rgb]{0,0.2,0.6}}
\usepackage[hmargin=.7in,vmargin=1.1in]{geometry}
\usepackage{indentfirst}
\newcommand{\FR}[2]{\displaystyle\frac{\,{#1}\,}{#2}}
\newcommand{\fr}[2]{\mbox{$\frac{\,{#1}\,}{#2}$}}
\newcommand{\n}{\nonumber}

\graphicspath{{fig/}}

\def\bge{\begin{equation}}
\def\ede{\end{equation}}
\def\bga{\begin{aligned}}
\def\eda{\end{aligned}}
\def\bgp{\begin{pmatrix}}
\def\edp{\end{pmatrix}}
\def\bgs{\begin{subequations}}
\def\eds{\end{subequations}}
\newcommand{\order}[1]{\mathcal{O}({#1})}
\def\di{{\mathrm{d}}}
\def\D{{\mathrm{D}}}
\def\Di{{\mathcal{D}}}

\def\O{\mathcal{O}}
\def\mb{\mathbf}

\def\pd{\partial}
\def\ld{{\mathscr{L}}}

\def\la{\langle}\def\ra{\rangle}
\def\sla{\slashed}

\def\tr{\mathrm{\,tr\,}}

\def\to{\rightarrow}
\def\To{\Rightarrow}
\def\ii{\mathrm{i}}

\def\al{\alpha}
\def\be{\beta}
\def\ga{\gamma}
\def\de{\delta}
\def\ep{\epsilon}
\def\ka{\kappa}
\def\lam{\lambda}
\def\rh{\rho}
\def\si{\sigma}

\newcommand{\ob}[1]{\mkern 2mu \overline{\mkern -2mu #1 \mkern -2mu}\mkern 2mu}
\newcommand{\wt}[1]{\mkern 2mu \widetilde{\mkern -2mu #1 \mkern -2mu}\mkern 2mu}

\begin{document}

\title{ \Large\textbf{Neutrino Signatures in Primordial Non-Gaussianities}}
\author{Xingang Chen$^{a}$\footnote{Email: xingang.chen@cfa.harvard.edu}{},~~ Yi Wang$^{b,c}$\footnote{Email: phyw@ust.hk}{},~ and~ Zhong-Zhi Xianyu$^{d,e}$\footnote{Email: xianyu@cmsa.fas.harvard.edu}\\[2mm]
\normalsize{$^a$\emph{Institute for Theory and Computation, Harvard-Smithsonian Center for Astrophysics,}}\\
\normalsize\emph{{60 Garden Street, Cambridge, MA 02138, USA}}\\
\normalsize{$^b$\emph{Department of Physics, The Hong Kong University of Science and Technology,}}\\
\normalsize{\emph{Clear Water Bay, Kowloon, Hong Kong, P.R.China}}\\
\normalsize{$^c$\emph{Jockey Club Institute for Advanced Study, The Hong Kong University of Science and Technology,}}\\
\normalsize{\emph{Clear Water Bay, Kowloon, Hong Kong, P.R.China}}\\
\normalsize{$^d$\emph{Center of Mathematical Sciences and Applications, Harvard University,}} \\
\normalsize{\emph{20 Garden Street, Cambridge, MA 02138, USA}}\\
\normalsize{$^e$~\emph{Department of Physics, Harvard University, 17 Oxford Street, Cambridge, MA 02138, USA}}}

 \date{}
\maketitle

\vspace{1cm}

\begin{abstract}

We study the cosmological collider phenomenology of neutrinos in an effective field theory. The mass spectrum of neutrinos and their characteristic oscillatory signatures in the squeezed limit bispectrum are computed. Both dS-covariant and slow-roll corrections are considered, so is the scenario of electroweak symmetry breaking during inflation. Interestingly, we show that the slow-roll background of the inflaton provides a chemical potential for the neutrino production. The chemical potential greatly amplifies the oscillatory signal and makes the signal observably large for heavy neutrinos without the need of fine tuning.

\end{abstract}

\newpage


\section{Introduction}
\label{sec_intro}

The inflationary universe provides us an invaluable opportunity of probing the physics at extremely high energies up to $\order{10^{14}\text{GeV}}$ through primordial fluctuations. These fluctuations have been measured through CMB and the large scale structure, and can be measured in unprecedented precision with 21~cm tomography in the future. The statistics of these fluctuations can be studied through the connected $n$-point ($n\geq 3$) correlation functions of primordial fluctuations, known as primordial non-Gaussianities. They recorded directly the information about the dynamics and the interactions of physics at the energy scale of inflation, and thus provided us a unique window to new physics beyond the well-established Standard Model (SM) of particle physics.

In this regard, an especially informative channel is the bispectrum (3-point function) of the scalar fluctuations in the squeezed limit, where the wave number of one external leg (long mode, ``long'' in terms of wavelength) is much smaller than the other two (short modes). It was shown in \cite{Chen:2009we,Chen:2009zp,Baumann:2011nk,Assassi:2012zq, Sefusatti:2012ye, Norena:2012yi, Chen:2012ge, Noumi:2012vr, Gong:2013sma, Emami:2013lma, Kehagias:2015jha, Arkani-Hamed:2015bza,  Dimastrogiovanni:2015pla,Chen:2016nrs, Lee:2016vti, Meerburg:2016zdz,Chen:2016uwp,Chen:2016hrz,
Kehagias:2017cym, An:2017hlx,Iyer:2017qzw,An:2017rwo, Kumar:2017ecc, Franciolini:2017ktv,Tong:2018tqf, MoradinezhadDizgah:2018ssw, Saito:2018xge,Franciolini:2018eno,Chen:2018sce,Saito:2018omt} that massive fields with mass close to the inflationary Hubble scale can generate characteristic momentum dependence, oscillatory or nonanalytical scaling,  in the squeezed bispectrum.  This effect arises from the nonlocal behavior of the massive fields and cannot be mimicked by any local effective operator. The utility of this nonlocal signal in the squeezed bispectrum was further generalized and emphasized in \cite{Arkani-Hamed:2015bza} and was dubbed the ``cosmological collider''.

A systematic study of ``cosmological collider phenomenology'' was initiated in \cite{Chen:2016nrs,Chen:2016uwp,Chen:2016hrz, Kumar:2017ecc} by carefully analysing the SM mass spectrum and signatures in the squeezed bispectrum. These SM background signals on the cosmological collider can be considered as a baseline for probing new physics on the cosmological collider. It was shown in \cite{Chen:2016nrs,Chen:2016uwp,Chen:2016hrz} that SM particles, typically those much lighter than the inflation energy scale, can acquire large mass corrections in inflation background, at least from the infrared-enhanced loop diagrams. These corrections can significantly distort the spectrum of SM fields. A good understanding of these distortions is necessary to identify the SM signatures, and also very helpful to distinguish different inflation scenarios. It was also shown in \cite{Chen:2016hrz} that the SM signatures in squeezed bispectrum enter firstly at 1-loop level if the electroweak symmetry is unbroken during inflation, and it was shown in \cite{Kumar:2017ecc} that tree-level signatures of SM are also possible if the electroweak symmetry is broken.

With SM signatures known, it is then apt to study similar signatures of new physics beyond SM. Among many expected pieces of new physics, the neutrino oscillation has arguably the most solid evidence. The most natural explanation of neutrino oscillation is nonzero masses of neutrinos introduced through a new sector of right-handed neutrinos. Being SM singlets, right-handed neutrinos can have Majorana mass and can be extremely heavy a priori. Through the mixing with the left-handed neutrinos via the Yukawa interaction with SM Higgs, a tiny mass is then introduced to the observed neutrinos, and this is the celebrated (Type-I) seesaw mechanism \cite{Minkowski:1977sc,GellMann:1980vs,Yanagida:1980xy,Mohapatra:1979ia}. In this scenario, the electroweak scale $\order{10^2\text{GeV}}$ is the geometric mean of the observed neutrino mass $\sim\order{\text{eV}}$ and the Majorana mass of the right-handed neutrinos, assuming $\order{1}$ Yukawa couplings. Therefore we see that a natural scale for right-handed neutrinos is around $\order{10^{13}\text{GeV}}$, which happens to be close to the target energy of the cosmological collider. Of course, both the the mass of right-handed neutrinos and the inflation scale can be much lower, but this still suggest that the cosmological collider could be an ideal machine of probing otherwise unreachable right-handed neutrinos. While a very massive right-handed neutrino can generate oscillatory signals in the squeezed bispectrum, similar oscillatory signals that we shall calculate in this paper can also be generated by any massive singlet fermion that couples to the inflaton. So the oscillatory signature does not belong exclusively to the right-handed neutrino that takes part in the seesaw mechanism, but more generally to any massive singlet fermion. However, we shall still use the term ``neutrino signature'' throughout the paper, since the term ``sterile neutrino'' is usually given to any massive singlet fermion in the context of cosmology.

There is yet another line of reasoning that motivates the study of neutrinos in inflation. As was pointed out in \cite{Chen:2011zf,Chen:2011tu,Chen:2012ja, Chen:2014joa,Chen:2014cwa,Chen:2015lza,Chen:2016cbe,Chen:2016qce}, the nonanalytical oscillatory behavior in the squeezed bispectrum generated by heavy fields also directly records the evolution of the scale factor $a(t)$ as a function of time in the primordial era. Therefore it is potentially a useful discriminator of different scenarios of primordial universe, which can be either inflationary or not. In this sense the heavy field is dubbed a ``standard clock'' of the primordial universe. For a scalar field, the nonanalytical behavior can be either monotonic scaling $(m\leq 3H/2)$ or oscillatory $(m>3H/2)$ depending on its mass $m$. Therefore, a heavy scalar $(m>3H/2)$ is required for the clock to tick. On the contrary, the behavior generated by a spinor field is always oscillatory, so a spinor field can possibly be a better clock than a scalar field. The right-handed neutrinos are possibly the best candidate for a fermionic standard clock, as all other fermions in SM remain massless during inflation so long as the electroweak symmetry is not broken.

In this paper, we shall study the spectrum of the neutrinos during inflation and their signatures in the correlation functions of curvature perturbation. The correction from the neutrino loops to the power spectrum is a scale-independent constant shift. This scale-invariant correction cannot be distinguished from other scale-invariant contributions. Due to the scale-invariance of the loop correction, the leading order effect of neutrinos arises in the shape dependence of the bispectrum, whose squeezed limit can be very informative as mentioned above.

Compared with previous studies for SM particles \cite{Chen:2016uwp,Chen:2016hrz}, the treatment here is greatly improved in that we shall include systematically the slow-roll (non-dS covariant) corrections \`a la effective field theory (EFT), while our previous studies have focused on dS covariant signatures only. An important lesson we draw from including slow-roll correction is that the mass and the spin cannot fully characterize the dispersion of a particle during inflation, due to the loss of full Poincar\'e (or dS) isometry. More parameters are needed, among which a non-unit sound speed is a well-known example. But more importantly than a non-unit sound speed, we shall show in this paper that the slow-roll background will introduce a nonzero chemical potential to the neutrino at the very leading order of inflaton-neutrino interactions. This chemical potential has important phenomenological consequence in that it can greatly enhance the signal strength in the non-Gaussianity even for very heavy intermediate field, which is otherwise highly suppressed by the Boltzmann factor. As a result, we shall show that the neutrino signatures, despite being loop-suppressed, can be naturally large in EFT without fine tuning.

The rest of the paper is organized as follows. In Sec.\;\ref{sec_overview}, we shall provide an overview of neutrinos in an inflationary background. We shall outline the effective interactions involving the neutrinos, the inflaton, and also the Higgs field. Then we summarize the basic kinematic properties of a neutrino during inflation, including its mode functions and propagators, firstly in dS background, and then including the slow-roll correction. This result also applies more broadly to any two-component Weyl spinor. Armed with these results, we then explore the inflationary correction to neutrino spectrum, and also the corresponding signatures in the squeezed bispectrum in various scenarios. We will firstly consider the dS covariant signatures in Sec.\;\ref{sec_dS}, and then include the slow-roll corrections with nonzero chemical potential in Sec.\;\ref{sec_sl}. We conclude in Sec.\;\ref{sec_concl}. We further review the derivation of the neutrino propagator in dS spacetime in App.\;\ref{app_dS}, of the neutrino mode functions with nonzero chemical potential in App.\;\ref{app_chp}, and of the Schwinger-Keldysh path integral for Weyl spinors in App.\;\ref{app_sk}.

\section{Neutrinos in an Inflationary Background}
\label{sec_overview}

We review in this section the basic kinematic and dynamic properties of neutrinos in an inflationary universe. To describe the interactions between the inflaton and the neutrinos, we write down all effective interactions up to dimension-7 operators involving the inflaton, the neutrinos, and the SM Higgs field. The SM Higgs is included here because the physical picture can change qualitatively if the electroweak symmetry is spontaneously broken by a nonzero Higgs vacuum expectation value (VEV). To preserve the approximate shift symmetry of the inflaton, we assume that the inflaton is derivatively coupled. Throughout the paper, we focus on one family of neutrinos only, and thus neglect flavor mixing for simplicity.

After evaluating these effective operators on the slow-roll inflaton background, the neutrino field will receive nontrivial correction in its quadratic terms in the Lagrangian. We shall then review the kinematic properties of neutrino field during inflation by analyzing its quadratic Lagrangian. We review the mode function and the propagator, on both dS background and on slow-roll background. On the dS background, we shall adopt a dS-covariant language which turns out to be very convenient. On the slow-roll background, however, we have to follow a space-time-asymmetric treatment.

Before entering the details, here we spell out some conventions and notations. It is always a good approximation to take the background metric to be exactly dS even when we consider the slow-roll corrections from interactions without dS symmetry. In the conformal coordinates, the metric reads
\bge
  g_{\mu\nu}=\FR{1}{(H\tau)^2}\big(-\di\tau^2+\di\mb x^2\big),
\ede
where $\tau\in(-\infty,0)$ is the conformal time and $H$ is the Hubble parameter, which is a constant in dS. Being maximally symmetric spacetime, the Riemann curvature has the simple form $R_{\mu\nu\rh\si}=H^2(g_{\mu\rh}g_{\nu\si}-g_{\mu\si}g_{\nu\rh})$.

The left-handed and right-handed neutrinos carries different SM charges and thus it is convenient to present everything in terms of two-component Weyl spinors in 4-dimensional spacetime. Our conventions for spinors will be mostly in line with \cite{Srednicki:2007qs}. See also \cite{Dreiner:2008tw} for a comprehensive review. In particular, we use undotted $\al,\be,\cdots$ and dotted $\dot \al,\dot \be,\cdots$ for left-handed and right-handed spinor indices, respectively. We also use Latin letters $m,n,\cdots$ for flat Lorentz indices and Greek indices $\mu,\nu,\cdots $ for curved spacetime indices. The undotted and dotted indices are contracted by $\ep^{\al\be}$ and $\ep^{\dot\al\dot\be}$, respectively, while the spin-1 objects contracting different types of indices are defined by $\si^m_{\al\dot\al}=(I,\vec\si)$ and $\bar\si^{m\dot\al\al}=(I,-\vec\si)$. It follows that $\bar\si^{m\dot \al\al}=\ep^{\al\be}\ep^{\dot\al\dot\be}\si^m_{\be\dot\be}$. $\si$'s with curved indices are defined via vierbein $e_\mu^n$, e.g., $\si_{\al\dot\al}^\mu\equiv e^\mu_m\si^m_{\al\dot\al}$. Finally, the usual gamma matrix is defined by
\bge
  \ga^\mu=\bgp 0 & \si^\mu_{\al\dot\al} \\ \bar\si^{\mu\dot\al\al} & 0 \edp.
\ede

\subsection{Effective Couplings between Neutrinos and the Inflaton}

In this subsection we review the interactions of neutrinos, both left-handed and right-handed, during inflation in a generic single field inflation model. For the neutrino to leave any imprints in the primordial non-Gaussianities, they have to couple to the inflaton field either directly or indirectly, similar to other SM fields considered in \cite{Chen:2016hrz}. In \cite{Chen:2016hrz}, only scalar-type couplings in the form of $f(X,\phi)\mathcal{O}_\text{SM}$ were considered, where $X\equiv (\pd_\mu\phi)^2$, $f$ is an arbitrary function, and $\mathcal{O}_\text{SM}$ is a scalar operator made from SM fields. If the shift symmetry $\phi\to \phi+\text{const.}$ is strictly respected, then $f$ depends only on $X$ but not directly on $\phi$. In this paper, we shall improve this parameterization under the principle of EFT, i.e., we parameterize the neutrino-inflaton couplings as effective operators made from all relevant fields and consistent with all symmetries, order by order in operator dimensions. Then, by assuming a universal cutoff scale $\Lambda$ and $\order{1}$ dimensionless coupling constants for all effective operators, only natural models are retained.

Before considering the neutrino-inflaton couplings, here we review the basics of neutrino sector of SM supplemented by right-handed neutrinos. In SM, three families of left-handed neutrinos $\nu_i$ ($i=e,\nu,\tau$) are upper components of $SU(2)_L$ lepton doublets $L_i=(\nu_i,\ell_i)^T$, while the right-handed neutrinos $N_i$ are SM singlets. The relevant terms in the Lagrangian are
\begin{align}
\label{neutrinoL}
\ld=\sqrt{-g}\bigg[&-\FR{1}{2}|\D_\mu\mb H|^2-V(\mb H)\n\\
&+L_i^\dag\ii \bar{\si}^\mu\D_\mu L_i+N_i^\dag\ii\bar{\si}^\mu\D_\mu N_i-\FR{1}{2}m_{N0}\big(N_iN_i+\text{h.c.}\big)\n\\&+\big(y_{ij}L_i^T\wt{\mb H} N_j +\text{h.c.}\big)\bigg],
\end{align}
where $\wt{\mb H}\equiv \ii\sigma_2\mb H^*$, $\mb H$ is the Higgs doublet in the SM, and we have assumed identical mass for all three right-handed neutrinos for simplicity.

From now on we will work with one family of neutrinos only for simplicity. The effective couplings among the neutrinos, the inflaton, and the SM Higgs, respecting all SM gauge symmetry and the shift symmetry of the inflaton, can be summarized as
\begin{align}
\label{OEFT}
\ld=\sqrt{-g}\bigg[\xi R\mb H^\dag\mb H+\sum_{i=1}^2 \FR{\lambda_{hi}}{\Lambda^{D_{hi}-4}}\O_{hi}+\sum_{i=1}^8 \FR{\lam_{ni}}{\Lambda^{D_{ni}-4}}\O_{ni}\bigg],
\end{align}
where $\mathcal{O}_{hi}$ denotes effective operators made of the Higgs and the inflaton, and $\mathcal{O}_{ni}$ denotes operators involving neutrinos and the inflaton, as well as Higgs. $D_{hi}$ and $D_{ni}$ denote the mass dimensions of $\mathcal{O}_{hi}$ and $\mathcal{O}_{ni}$, respectively, while $\lam_{hi}$ and $\lam_{ni}$ are dimensionless constants which we assume to be of $\order{1}$. We also include the dim-4 non-minimal coupling between the Ricci scalar and the Higgs as it could be important in determining the background value of the Higgs. For $\mathcal{O}_{hi}$, we consider following two operators of dim-5 and dim-6, respectively,
\begin{align}
\label{Oh}
   &\O_{h1}=(\pd_\mu\phi)\mb H^\dag\pd^\mu\mb H,
  &&\O_{h2}=(\pd_\mu\phi)^2\mb H^\dag\mb H.
\end{align}
For $\O_{ni}$, we include following operators up to dim-7, where it is understood that complex operators are to be supplemented by their complex conjugate.
\bgs
\label{On}
\begin{align}
   &\text{dim-5:}
  &&\O_{n1}=L^\dag (\ob{\slashed{\pd}}\phi) L,
  &&\O_{n2}=N^\dag (\ob{\slashed{\pd}}\phi) N,\\
   &\text{dim-6:}
  &&\O_{n3}=(\square \phi)NN,
  &&\O_{n4}=\wt{\mb H}L^\dag(\ob{\sla{\pd}}\phi)N,\\
   &\text{dim-7:}
  &&\O_{n5}=(\pd_\mu\phi)^2NN,
  &&\O_{n6}=(\square\phi)L^\dag\ii\ob{\sla\D}L,\n\\
   &{}
  &&\O_{n7}=(\square\phi)N^\dag\ii\ob{\sla\D}N,
  &&\O_{n8}=(\square\phi)\wt{\mb H}L  N,\n\\
   &{}
  &&\O_{n9}=(\pd^\mu\phi)\wt{\mb H}L \D_\mu N,
  &&\O_{n10}=\mb H^\dag\mb H L^\dag (\ob{\slashed{\pd}}\phi) L,\n\\
   &{}
  &&\O_{n11}=\mb H^\dag\mb H N^\dag (\ob{\slashed{\pd}}\phi) N.
\end{align}
\eds

Many operators here can contribute to the mass of neutrinos after evaluated on the inflation background $\la\pd_\mu\phi\ra=\de_{\mu 0}\dot\phi_0$. In addition, the Higgs field may also develop a nonzero VEV during inflation $\la \mb H\ra = (0,v/\sqrt{2})^T$. Here we have rotated the nonzero component to the lower real direction. We stress that the VEV $v$ is independent of its low energy value $246$GeV. In subsequent sections, we shall study the spectrum of neutrinos, and we distinguish the symmetric phase $v=0$ and symmetry-broken phase $v\neq 0$. In addition, we also consider a possibility that the inflaton couples to SM only via the Higgs portal coupling, i.e. all couplings but $\xi$ and $\lambda_{h2}$ are set to zero in (\ref{OEFT}). But before elaborating those points, it is helpful to summarize the kinematic properties of a neutrino in inflation, include its mode functions and propagators, as we shall do now in the next two subsections.

\subsection{Weyl Spinors on dS Background}

Now we review the propagator for a Weyl spinor in exact dS spacetime, using the fully dS-covariant language, similar to the treatment in \cite{Chen:2016hrz}. As we shall see below, this language provides a neat way for calculating the dS-covariant loop corrections. The formalism presented here applies to any two-component Weyl spinor in dS spacetime, and not only to neutrinos. Therefore, we begin   with the Lagrangian of a two-component left-handed spinor $\psi_\al$ of mass $m$,
\bge
\ld=\sqrt{-g}\Big(\ii\psi^\dag\bar\si^\mu\D_\mu\psi-\FR{1}{2}m\psi\psi-\FR{1}{2}m\psi^\dag\psi^\dag\Big),
\ede
where $\D_\mu$ is the covariant derivative of the spinor in curved spacetime. We always denote the right-handed spinor with a dagger to avoid confusion. The equations of motion for the spinors are
\begin{align}
&\ii\si^\mu_{\al\dot \al}\D_\mu\psi^{\dag\dot\al}=m\psi_\al,\\
&\ii\bar\si^{\mu\dot \al\al}\D_\mu\psi_\al=m\psi^{\dag\dot\al}.
\end{align}
There are two types of propagators, $\la\psi^{\dag\dot \al}(x)\psi_{\al'}(x')\ra$ and $\la\psi_\al(x)\psi_{\al'}(x')\ra$. They satisfy the following equations,
\begin{align}
\label{GreenEqn1}
\ii\si^\mu_{\al\dot \al}\D_\mu\la\psi^{\dag\dot\al}(x)\psi_{\al'}(x')\ra=&~m\la\psi_\al(x)\psi_{\al'}(x')\ra,\\
\label{GreenEqn2}
\ii\bar\si^{\mu\dot\al\al}\D_\mu\la\psi_\al(x)\psi_{\al'}(x')\ra=&~m\la\psi^{\dag\dot\al}(x)\psi_{\al'}(x')\ra.
\end{align}
We parameterize these two propagators according to their symmetry structures by
\begin{align}
\label{spinorprop1}
\la\psi^{\dag\dot \al}(x)\psi_{\al'}(x')\ra=&~f(Z)P_{\be\al'}n_\mu\bar\si^{\mu\dot\al\be},\\
\label{spinorprop2}
\la\psi_\al(x)\psi_{\al'}(x')\ra=&~g(Z)P_{\al\al'},
\end{align}
where $f(Z)$ and $g(Z)$ are scalar functions of the dimensionless imbedding distance $Z$ between $x$ and $x'$, which is related to the geodesic distance $L$ via $Z=\cos(HL)$. $P_{\al\al'}$ is the parallel translator of left-handed spinor, which is itself a bispinor and is defined by $n^\mu\D_\mu P_\al{}^{\al'}=0$. $n_\mu=\D_\mu L$ is the unit vector at $x$ tangent to the geodesic between $x$ and $x'$. In exact dS background, it is possible to solve the equation of propagators (\ref{GreenEqn1}) and (\ref{GreenEqn2}) directly, without referring to mode functions. This was derived in \cite{Allen:1986qj} and we review the derivations in App.\;\ref{app_dS}. Here we summarize the solutions for the scalar functions  $f(Z)$ and $g(Z)$ as
\begin{align}
  f(Z)=&-\FR{\ii H^3\Gamma(2-\ii m/H)\Gamma(2+\ii m/H)}{16\sqrt{2}\pi^2}\sqrt{1-Z}{\,}_2F_1\Big(2-\FR{\ii m}{H},2+\FR{\ii m}{H};2;\FR{1+Z}{2}\Big),\\
  g(Z)=&~\FR{H^3\Gamma(2-\ii m/H)\Gamma(2+\ii m/H)}{32\sqrt 2\pi^2}\FR{m}{H}\sqrt{1+Z}{\,}_2F_1\Big(2-\FR{\ii m}{H},2+\FR{\ii m}{H};3;\FR{1+Z}{2}\Big).
\end{align}

\subsection{Weyl Spinors on Slow-Roll Background}

After coupling a fermion to the inflaton field in the effective field theory approach, the fermion acquires non-dS invariant terms due to the slow-roll background of the inflaton $\dot\phi_0\neq 0$. The dS covariant language presented above no longer applies straightforwardly, and thus it is useful to work out the mode function of a fermion including possible non-dS covariant terms.

We will begin with the following Lagrangian, which is the most general Lagrangian for neutrinos in EW-symmetric phase after including inflaton coupling to dim-7 operators. The reason that we consider dim-7 operator is that this is the lowest order at which the neutrino mass receives corrections from the inflaton background.
\begin{align}
\label{quardldcp}
  \ld = a^3 \psi^\dag\ii \ob{\sla{\D}}\psi- a^4\lam \psi^\dag\ob\si^0\psi-\FR{1}{2}a^4m(\psi\psi+\text{c.c.}),
\end{align}
where all Lorentz indices should be lowered or raised by the flat Minkowski metric $\eta_{mn}$ or its inverse. Here we observe that the spectrum of the neutrino field is characterized by two parameters, the ordinary Majorana mass $m$ and a new coupling $\lam$ from a dim-5 inflaton coupling, which also has the dimension of the mass. We redefine $\psi(x)=a^{-3/2}\wt\psi(x)$ as usual so that the covariant derivative on $\psi$ simplifies to the partial derivative on $\wt\psi$. In terms of $\wt\psi$, the above Lagrangian becomes
\begin{align}
  \ld = \wt\psi^\dag\ii \ob{\sla{\pd}}\wt\psi- a \lam \wt\psi^\dag\ob\si^0\wt\psi-\FR{1}{2}a m(\wt\psi\wt\psi+\text{c.c.}).
\end{align}
The equations of motion can be obtained by varying the Lagrangian with $\wt\psi^\dag$ and the result is
\begin{align}
\label{eqomcp}
  \ii \bar\si^{m\dot\al\be}\pd_m\wt\psi_\be =a\lam\bar\si^{0\dot\al\be}\wt\psi_\be+am\wt\psi^{\dag\dot \al}.
\end{align}
To proceed, we decompose the spinor $\wt\psi$ in terms of eigenmodes of 3-momentum,
\bge
\label{mode}
  \wt\psi_\al(\tau,\mb x)= \int\FR{\di^3\mb k}{(2\pi)^3}\sum_{s=\pm}\Big[\xi_{\al,s}(\tau,\mb k)b_s(\mb k)e^{+\ii\mb k\cdot\mb x}+\chi_{\al,s}(\tau,\mb k)b_s^{ \dag}(\mb k)e^{-\ii\mb k\cdot\mb x}\Big],
\ede
where $s$ is the helicity index, $b_s$ and $b_s^\dag$ are annihilation and creation operators satisfying the usual anticommutation relation $[b_s(\mb k),b_{s'}(\mb k')]_+=(2\pi)^3\de_{ss'}\de^{(3)}(\mb k-\mb k')$, and the coefficients $\xi_{\al,s}(\mb k)$ and $\chi_{\al,s}(\mb k)$ are coefficients with definite helicity. We rewrite these two spinors in terms of unit helicity eigenstates $h_s$,
\begin{align}
\label{helimode}
  &\xi_{\al,s}(\tau,\mb k)=u_s(\tau,\mb k)h_s(\mb k), && \chi^{\dag\dot \al}_s(\tau,\mb k) = v_s(\tau,\mb k) h_s(\mb k).
\end{align}
We choose $\chi^\dag$ instead of $\chi$ in the above expression so that both $u$ and $v$ correspond to modes with positive frequency. The unit helicity eigenspinor $h_s$ satisfies the following relations.
\begin{align}
  &\vec\si\cdot\vec k h_s(\mb k) = s k\,h_s(\mb k),
  &&h^\dag_s(\mb k) h_{s'}(\mb k) = \de_{ss'},
  &&\sum_{s=\pm}h_s(\mb k) h_s^\dag(\mb k) = 1.
\end{align}
Substituting the mode expansion (\ref{mode}) back into the equation of motion (\ref{eqomcp}), we can find the properly normalized solutions to the mode functions $u_s$ and $v_s$ as
\begin{align}
\label{uplus}
  u_{+}(\tau,\mb k)=&~\FR{\wt m e^{+\pi\wt\lam/2}}{\sqrt{-2k\tau}}W_{\ka,\ii\wt\mu}(2\ii k\tau) ,\\
\label{uminus}
  u_{-}(\tau,\mb k)=&~\FR{e^{-\pi\wt\lam/2}}{\sqrt{-2k\tau}}W_{-\ka,\ii\wt\mu}(2\ii k\tau) ,\\
\label{vplus}
  v_+(\tau,\mb k)=&~\FR{e^{+\pi\wt\lam/2}}{\sqrt{-2k\tau}}W_{1+\ka,\ii\wt\mu}(2\ii k\tau) ,\\
\label{vminus}
  v_-(\tau,\mb k)=&~\FR{\wt m e^{-\pi\wt\lam/2}}{\sqrt{-2k\tau}}W_{-1-\ka,\ii\wt\mu}(2\ii k\tau),
\end{align}
where $W_{\ka,\mu}(z)$ is the Whittaker function, and
\begin{align}
\label{lammu}
  &\ka=-\FR{1}{2}-\ii\wt\lam,
  &&\wt m=\FR{m}{H},
  &&\wt\lam=\FR{\lam}{H},
  &&\wt\mu=\sqrt{\wt m^2+\wt\lam^2}.
\end{align}
More details about deriving these solutions are collected in App.\;\ref{app_chp}.

When extracting the oscillation signals in the squeezed bispectrum, we are mostly interested in the late-time behavior of these mode functions. Therefore we show here the leading terms in $|k\tau|\ll 1$ limit,
\bgs
\label{mfltl}
\begin{align}
 u_+(\tau,\mb k)\simeq &~ e^{-\ii\pi/4}e^{+\pi\wt\lam/2}\wt m\bigg[\FR{e^{\pi\wt\mu/2}\Gamma(-2\ii\wt\mu)}{\Gamma(1+\ii\wt\lam-\ii\wt\mu)}(-2k\tau)^{\ii\wt\mu}+(\wt\mu\to-\wt\mu)\bigg],\\
 u_-(\tau,\mb k)\simeq &~ e^{-\ii\pi/4}e^{-\pi\wt\lam/2} \bigg[\FR{e^{\pi\wt\mu/2}\Gamma(-2\ii\wt\mu)}{\Gamma(-\ii\wt\lam-\ii\wt\mu)}(-2k\tau)^{\ii\wt\mu}+(\wt\mu\to-\wt\mu)\bigg],\\
 v_+(\tau,\mb k)\simeq &~ e^{-\ii\pi/4}e^{+\pi\wt\lam/2} \bigg[\FR{e^{\pi\wt\mu/2}\Gamma(-2\ii\wt\mu)}{\Gamma(\ii\wt\lam-\ii\wt\mu)}(-2k\tau)^{\ii\wt\mu}+(\wt\mu\to-\wt\mu)\bigg],\\
 v_-(\tau,\mb k)\simeq &~ e^{-\ii\pi/4}e^{-\pi\wt\lam/2}\wt m\bigg[\FR{e^{\pi\wt\mu/2}\Gamma(-2\ii\wt\mu)}{\Gamma(1-\ii\wt\lam-\ii\wt\mu)}(-2k\tau)^{\ii\wt\mu}+(\wt\mu\to-\wt\mu)\bigg].
\end{align}
\eds
As we shall show below, the oscillatory dependence in time $\propto \tau^{\pm\ii\wt\mu}$ will eventually be translated to the oscillatory signal as a function of the momentum ratio in the squeezed bispectrum. Since $\wt\mu$ here is always real, the signal is always oscillatory, as mentioned in Sec.\;\ref{sec_intro}.

The most important feature of the late-time limit (\ref{mfltl}) is the effect of a nonzero $\lambda$, which can be interpreted as a chemical potential. The consequence is two-fold. Firstly, a nonzero $\lambda$ enters the time dependence through the combination $\tau^{\pm\ii\sqrt{m^2+\lam^2}/H}$. Compared with the case of vanishing $\lambda$, we see that $\lam$ correct the apparent value of the fermion mass in the oscillatory signal. So in this sense we may write $m_\text{clock}^2=m^2+\lam^2$.

To see the second effect of $\lambda$ which is the most important one, we derive the nonanalytical part of the fermion propagator in the late-time limit. The full expressions for Weyl fermion propagators in the in-in formalism are summarized in App.\;\ref{app_sk}. There are in total eight different propagators for a Weyl spinor in the in-in formalism, depending on both spinor indices and in-in indices. For our purpose here it suffices to show one example,
\begin{align}
\label{proplatetime}
  D_{-+\al\dot\be}
  =&~\xi_\al(\tau_1,\mb k)\xi_{\dot\be}^\dag (\tau_2,\mb k)\n\\
  =&~\bigg\{\bigg[\FR{-e^{+\pi\wt\lam}\Gamma^2(-2\ii\wt\mu)}{\Gamma(\ii\wt\lam-\ii\wt\mu)\Gamma(-\ii\wt\lam-\ii\wt\mu)}h_+(\mb k)h_+^\dag(\mb k) + \FR{\wt m\Gamma^2(-2\ii\wt\mu)}{\Gamma(1+\ii\wt\lam-\ii\wt\mu)\Gamma(\ii\wt\lam-\ii\wt\mu)} h_+(\mb k)h_-^\dag(\mb k) \n\\
  &~+ \FR{\wt m\Gamma^2(-2\ii\wt\mu)}{\Gamma(1-\ii\wt\lam-\ii\wt\mu)\Gamma(-\ii\wt\lam-\ii\wt\mu)} h_-(\mb k)h_+^\dag(\mb k) +
  \FR{e^{-\pi\wt\lam}\Gamma^2(-2\ii\wt\mu)}{\Gamma(\ii\wt\lam-\ii\wt\mu)\Gamma(-\ii\wt\lam-\ii\wt\mu)} h_-(\mb k)h_-^\dag(\mb k)\bigg]\n\\
  &~\times(4k^2\tau_1\tau_2)^{+\ii\wt\mu}\bigg\} +~(\wt\mu\to-\wt\mu) ~.
\end{align}

Here we only keep the nonanalytical part of the late-time expansion with imaginary power dependence on the time. From this expression, we can see that $\wt\lam$ corrects the coefficient of $(\tau_1\tau_2)^{\pm \ii\wt\mu}$ in a way that cannot be described by a mass shift $m^2\to m^2+\lam^2$. To appreciate the importance of this effect, we consider the large mass limit of the mode function, using the approximation $\Gamma(x+\ii y)\sim e^{-\pi|y|/2}$ when $y\to \infty$, with $x$ and $y$ both real. Then, in the absence of $\lam$, it can be readily seen from (\ref{proplatetime}) that  the large mass limit $\wt m\gg 1$ leads to the familiar Boltzmann suppression $\sim e^{\pi \wt m }$ for all helicity components. On the contrary, if we allow a large chemical potential $\lam$, then, in the limit of $\wt\lam\gg 1$ and $\wt\lam\gg \wt m$, we see that only $h_-h_-^\dag$ component has exponential suppression $\propto e^{-2\pi\wt\lam}$ while all other three components are free of exponential suppression. Similarly, if we choose $\wt\lam<0$ while $|\wt\lam|\gg \wt m$, then only the positive-helicity part $h_+^{}h_+^\dag$ will be exponentially suppressed. Physically this is because a positive chemical potential $\lam$ will boost the production of fermions with positive helicity, and suppress the particle number with negative helicity. Equivalently, a negative chemical potential will boost the production of fermions with positive helicity. Either way, there will be a component in the Weyl spinor with definite helicity, whose Boltzmann suppression from large $m$ will be delayed by a nonzero chemical potential. In Sec.\;\ref{sec_sl}, we shall show that this delay of Boltzmann suppression will allow a naturally large contribution to the oscillatory signals in the squeezed limit, even at the 1-loop level. We note that similar effect of this chemical potential were considered in \cite{Adshead:2015kza,Adshead:2018oaa} in the context of fermion production.

\section{dS Covariant Signatures of Neutrinos}
\label{sec_dS}

In this section, we study the correction to the mass spectrum of the neutrinos and their oscillatory signatures in the squeezed bispectrum. In parallel with the analysis in \cite{Chen:2016hrz}, we consider in this section only the dS covariant corrections, where the mass and spin suffice to describe the kinematic properties of a particle. This is by no means the most general situation, but we choose to concentrate on this case here and to include slow-roll corrections only in next two sections, because the calculation in this case is greatly simplified by using the dS covariant formalism, as outlined in \cite{Chen:2016hrz}, and summarized in Sec.\;\ref{sec_overview} in this paper. See also App.\;\ref{app_dS} for more details. In addition, we assume that the electroweak is not broken during inflation in this section, and consider the broken case in the next section. As we shall see in the next section, if the inflaton field $\phi$ is subject to a reflective $\mathbb{Z}_2$ symmetry, $\phi\to -\phi$, the result of this section will also apply even when general slow-roll corrections are included in the EFT fashion.

\subsection{dS Covariant Mass Correction to Neutrinos from the Higgs Loop}

In general, the mass of neutrinos, both left-handed and right-handed, receives nonzero corrections during inflation. At the very least, the Higgs can contribute a nonzero mass to neutrinos through the following 1-loop process,
\bge
\label{mass1loopdiag}
  \parbox{0.4\textwidth}{\includegraphics[width=0.4\textwidth]{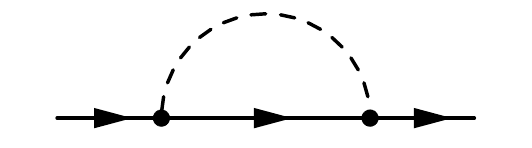}}
\ede
where the arrowed lines are fermion propagator and the dashed line is the Higgs propagator. This correction is present even when the electroweak symmetry is unbroken, as detailed in \cite{Chen:2016hrz}. The key point to be noted is that the propagator of a nearly massless scalar field in dS receives an infrared enhancement inversely proportional to the mass square in the limit $m/H\ll 1$. This fact is mostly easily seen from the Wick-rotated dS, namely a sphere $S^4$, on which a scalar field can be decomposed into spherical harmonics $Y_{\vec{L}}(x)$ where $\vec L=(L,L_3,L_2,L_1)$ denotes collectively all indices of the 4-dimensional spherical harmonic function, which are all integers and satisfy $L\geq L_3\geq L_2\geq |L_1|$. Then, the propagator for a scalar field $\phi$ with mass $m$ can be worked out to be \cite{Chen:2016hrz}
\bge
\label{scalarprop}
  \la\phi(x)\phi(y)\ra = \sum_{\vec L}\FR{H^2}{L(L+3)+(m/H)^2}Y_{\vec{L}}^{}(x)Y_{\vec{L}}^* (y).
\ede
The zero mode corresponds to $L=0$ component, and it is clear that the propagator is dominated by the zero mode and behaves like $1/m^2$ when $m\to 0$.

Consequently, if the scalar mass satisfies $m\ll H$, then any loop process involving a scalar zero-mode running in the loop would receive corresponding enhancement, and it is in general a very good approximation of the loop result to keep the zero-mode contribution only. On the other hand, if the scalar is heavy, $m\gg H$, then the loop correction from this scalar field would be negligibly small. In \cite{Chen:2016hrz}, we show that the Higgs mass during inflation is generally $\order{\lam_h^{1/4}H}$ with $\lam_h$ is strength of the Higgs self-coupling, if there is no mass correction from the inflation background. After including the latter, the Higgs mass would be different, but would in general be of order $H$. Therefore, when consider the Higgs-loop correction to the neutrino mass, we shall work with the approximation of zero-mode domination.

With the above strategy in mind, we now consider the mass correction to neutrinos from the Higgs-loop. The Yukawa interaction between the Higgs and neutrinos is given in the last line of (\ref{neutrinoL}) which we iterate for one family of neutrinos,
\bge
  \Delta\ld = \sqrt{-g} y h (\nu N+ \nu^\dag N^\dag),
\ede
where $y$ is the Yukawa coupling, $h$ is the CP-even neutral component of the Higgs field. We note that both the ``left-handed'' $\nu$ and ``right-handed'' $N$ neutrinos are actually left-handed Weyl spinors, while the right-handed spinors are always marked with $\dag$'s. Furthermore, if the electroweak symmetry is unbroken, we know that, at tree level, the left-handed neutrinos remain massless, $m_{\nu 0} =0 $, the right-handed neutrino has a tree level mass which we denote by $m_{N0}$, and the Higgs field $h$ can have an arbitrary mass $m_h$ which can be either smaller or greater than the Hubble scale.

\begin{figure}[tbph]
\centering
\includegraphics[width=0.55\textwidth]{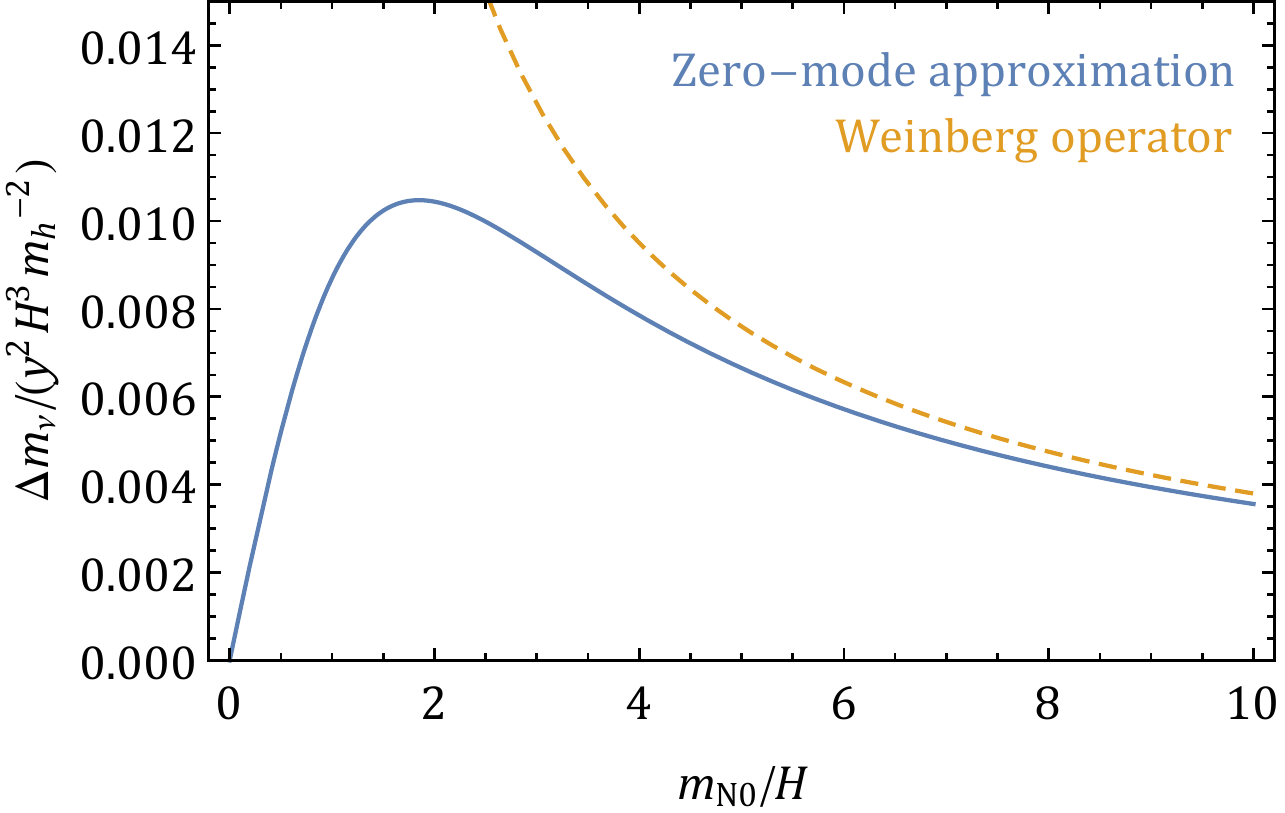}
\caption{1-loop corrections to the mass of the left-handed neutrino $\Delta m_\nu$ as functions of right-handed neutrino mass $m_{N0}$. The blue solid curve shows the correction from Yukawa interaction derived in (\ref{Dmnu}) with zero-mode approximation, while the dashed orange curve shows the correction from the effective Weinberg operator valid in large $m_{N0}$ limit.}
\label{FigDeltaM}
\end{figure}

Since we are interested in the mass correction from the 1-loop diagram, we can set the external momentum of the neutrino to zero. Then the loop correction to the mass of the left-handed neutrino $\nu$ from the diagram in (\ref{mass1loopdiag}) has the following form
\bge
\label{1loopint}
  \Delta m_\nu P_{\al\al'}\int\di\Omega=y^2\int\di\Omega\di\Omega'\,\la h(x) h(x')\ra\,\la N_\al(x)N_{\al'}(x')\ra,
\ede
where $\di\Omega$ denotes the dS-invariant integral measure on $S^4$, the Higgs propagator $\la h(x)h(x')\ra$ is given in (\ref{scalarprop}) with the mass $m=m_h$, and the neutrino propagator $\la N_\al(x)N_{\al'}(x')\ra$ given by (\ref{spinorprop2}). The approximation of zero-mode dominance means that we can substitute the Higgs propagator by its zero-mode component, which is a constant and is given by ${3H^4}/(8\pi^2m_h^2)$. Moreover, since there is no ultraviolet divergence associated with zero modes, we can simply work in 4 dimensions. Then we have $\la N_\al(x)N_{\al'}(x')\ra$ in the integrand only. We then take away one layer of integral $\di\Omega'$ since the propagator depends only on imbedding distance $Z(x,x')$. Consequently, the above expression (\ref{1loopint}) for the mass correction can be rewritten as
\begin{align}
\label{MF1loop}
&~\Delta m_\nu P_{\al\al'}=\FR{3y^2H^3}{8\pi^2 m_h^2}\int\di\Omega\,\la N_\al(x)N_{\al'}(x')\ra
=\FR{3y^2H^4}{8\pi^2m_h^2}V_{3}P_{\al\al'}\int_{-1}^1\di Z\,(1-Z^2) g(Z)
\end{align}
In the second equality of above expression, we have finished the integration over $S^3$, yielding the volume factor $V_3=2\pi^2H^{-3}$. The final integral can be carried out analytically, and the mass correction can thus be found to be
\bge
\label{Dmnu}
 m_{\nu}=\FR{3y^2H^2 m_{N0}\Gamma(2-\ii m_{N0}/H)\Gamma(2+\ii m_{N0}/H)}{140\pi^2 m_h^2}{}_3F_2\Big(2-\FR{\ii m_{N0}}{H},2+\FR{\ii m_{N0}}{H},\FR{5}{2};3,\FR{9}{2};1\Big),
\ede
in which we have written $\Delta m_\nu$ simply as $m_\nu$ since the tree-level value of $m_\nu$ is zero. The ${}_3F_2$ factor is finite at $m_{N0}=0$. Thus we see that the above mass correction vanishes when $m_{N0}=0$, which recovers the result found in \cite{Chen:2016hrz}. One can also understand the vanishing result of $m_{N0}\to 0$ limit by noting that the propagator $\la N_\al(x)N_{\al'}(x')\ra$ itself is zero when $m_{N0}=0$. On the other hand, when $m_{N0}/H\gg 1$, the above expression behaves like
\bge
\label{WO}
  m_\nu \sim \FR{3y^2H^4}{8\pi^2m_h^2m_{N0}}=\FR{y^2}{m_{N0}}\la h^2\ra,
\ede
as one would expect from the dimension-5 Weinberg operator $LL\mb H\mb H $. We plot (\ref{Dmnu}) and (\ref{WO}) in Fig.\;\ref{FigDeltaM} from which the behavior of the mass correction in the two limits can be clearly seen.

Above we found the mass correction to the left-handed neutrino from the loop formed with one Higgs line and one right-handed neutrino line. The result vanishes when the neutrino mass in the loop is sent to zero. From this we immediately know that, since the left-handed neutrino $\nu$ is massless at tree level, its mass correction to the right-handed neutrino is also zero. One may worry that $\nu$ may contribute a mass to $N$ after receiving $\Delta m_\nu$ as shown in (\ref{Dmnu}). This contribution, however, can be neglected since $\Delta m_{\nu}\ll m_{N0}$ so long as $m_h$ is not extremely smaller than $H$, which is expected to be true, so we shall not consider the back reaction any further.

Now we get a diagonal mass matrix for both left- and right-handed neutrinos,
\bge
  \ld\supset-\FR{1}{2}\sqrt{g}\bgp \nu & N \edp\bgp \Delta m_\nu & 0 \\ 0 & m_{N0} \edp \bgp \nu \\ N \edp +\text{h.c.}.
\ede
We see that the mass generation here is quite different from the seesaw mechanism that works in the broken phase of electroweak symmetry.

\subsection{dS Covariant Signatures in Bispectrum}

In the previous subsection we see that the left-handed neutrino receives nonzero mass correction from the infrared enhanced 1-loop of Higgs, even when the electroweak symmetry is unbroken. The right-handed neutrino, on the other hand, is in general massive, too. The question then is to find the signatures of these massive spinors in the primordial non-Gaussianity. As described in Sec.\;\ref{sec_intro}, the most distinct signature of such massive spinor fields is the one from the squeezed limit of the bispectrum, where the neutrinos leave their imprints through their coupling to the inflaton. Since fermions have to be created in pairs, the leading order contribution would again be a 1-loop process. 
 The characteristic oscillatory signals, then, come from the region where the two internal neutrino propagators are attached to the soft external leg of the squeezed configuration, so that they become the clock field for the shorter curvature modes when these soft modes enter the classical regime \cite{Chen:2015lza}. 
In this region, we are allowed to expand the internal momentum in the late-time limit, as will be detailed in the following.

In this subsection, we will again focus on a dS-covariant signature, arising from a general interaction between the inflaton and a two-component Weyl spinor of the following form,
\bge
\label{dScovint}
\ld\supset-\FR{1}{2}\sqrt{-g}f_\psi(X)(\psi\psi+\psi^\dag\psi^\dag),
\ede
where the Weyl spinor $\psi$ of mass $m$ represents either left- or right-handed neutrino, $X\equiv(\pd_\mu \phi)^2$, and $f_\psi(z)$ is an arbitrary function. In the next section, we will find a more specific form of $f_\psi(z)$ in the EFT framework. We note in passing that this operator will also contribute to the neutrino mass during inflation, $\Delta m_\psi = f_\psi(X_0)$, after evaluated on the inflaton background $\la X\ra= X_0 =-\dot\phi_0^2$. We shall consider this part of correction more carefully in the next section, too. Furthermore, we assume the neutrino field $\psi$ here has canonically normalized kinetic term, so that its propagator is still given by (\ref{spinorprop1}) and (\ref{spinorprop2}).

The 1-loop diagram that contribute to the oscillatory signal of the squeezed bispectrum is
\bge
\label{ds3ptdiag}
  \parbox{0.4\textwidth}{\includegraphics[width=0.36\textwidth]{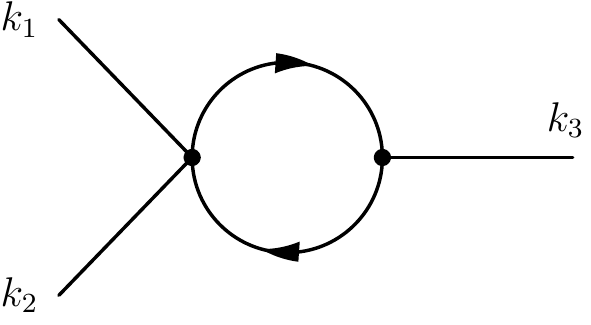}}
\ede
The three external lines represent propagators of the inflaton perturbation $\de\phi$ with momenta $k_i~(i=1,2,3)$. We take the squeezed limit $k_3\ll k_{1,2}$. The four-point vertex and three-point vertex in the above diagram come from expanding the interaction (\ref{dScovint}) around the inflaton background $X=X_0$. Therefore, the four-point interaction is
\bge
  -\FR{1}{2}\sqrt{-g}f_\psi(X_0)(\pd_\mu\de\phi)^2(\psi\psi+\psi^\dag\psi^\dag),
\ede
and the three-point vertex is
\bge
  \sqrt{-g}f_\psi(X_0)\dot\phi_0\dot{\de\phi}(\psi\psi+\psi^\dag\psi^\dag).
\ede
Then it is straightforward to write down the following expression for the neutrino loop in (\ref{ds3ptdiag}),
\begin{align}
&\FR{1}{4}\Big\la\big[\psi^\al(x)\psi_\al(x)+\psi_{\dot\al}^\dag(x)\psi^{\dag\dot\al}(x)\big]
\big[\psi^{\be'}(x')\psi_{\be'}(x')+\psi_{\dot \be'}^\dag(x')\psi^{\dag\dot \be'}(x')\big]\Big\ra\n\\
=&-\FR{1}{2}\Big[\ep^{\al\ga}\ep^{\be'\de'}\la \psi_\al\psi_{\be'}\ra \la \psi_\ga \psi_{\de'}\ra+\ep_{\dot \al\dot \ga}\ep^{\be'\de'}\la\psi^{\dag\dot\al}\psi_{\be'}\ra \la \psi^{\dag\dot\ga}\psi_{\de'}\ra+\text{h.c.}\Big]\n\\
=&-\FR{1}{2}\Big[g^2(Z)+f^2(Z)\Big]P_{\al\be'}P^{\al\be'}+\text{h.c.}=-2\Big[g^2(Z)+f^2(Z)\Big],
\end{align}
in which we have substitute the fermion propagators, (\ref{spinorprop1}) and (\ref{spinorprop2}), and also used the property $P_{\al\be'}P^{\al\be'}=2$ and the fact that $f^2$ and $g^2$ are manifestly real. Now it is straightforward to expand $f(Z)$ and $g(Z)$ at late time limit $\tau\to 0$, and keep the nonlocal part only,
\begin{align}
-2\big[f^2(Z)+g^2(Z)\big]=-\FR{3H^6}{8\pi^5}(1-2\ii \wt m)\Gamma^2(2-\ii \wt m)\Gamma^2(-\FR{1}{2}+\ii\wt m)\Big(\FR{\tau\tau'}{X^2}\Big)^{4-2\ii \wt m}+\text{c.c.},
\end{align}
where $\wt m\equiv m/H$. With this expression for the loop, we can then calculate the 3-point function of the inflaton perturbation following the standard in-in formalism. The details of this method can be found in \cite{Chen:2016hrz} and here we only show the result,
\begin{align}
\la\de\phi(\mb k_1)\de\phi(\mb k_2)\de\phi(\mb k_3)\ra'=-\FR{3 f_\psi^2(X_0)H^7\dot\phi_0}{2\pi^4k_1^6}C_\psi(\wt m)\Big(\FR{k_3}{2k_1}\Big)^{1-2\ii \wt m}+\text{c.c.},
\end{align}
where $\la\cdots\ra'$ represents the correlation function with the $\de$-function of momentum conservation stripped off, and the function $C_\psi(\wt m)$ is
\begin{align}
C_\psi(\wt m)\equiv&~ \ii(2-\ii \wt m)(1-2\ii \wt m)(5-2\ii \wt m)\Gamma^2(2-\ii \wt m)\Gamma^2(-\FR{1}{2}+\ii \wt m)\n\\
&~\times\Gamma^2(3-2\ii \wt m)\Gamma(-6+4\ii \wt m)\cosh^3(\pi \wt m)\sinh(\pi \wt m).
\end{align}
It is conventional to represent the non-Gaussianity in terms of the dimensionless shape function $S$, defined from the following expression,
\bge\label{3ptshaperaw}
\langle \zeta(\mathbf{k}_1) \zeta(\mathbf{k}_2) \zeta(\mathbf{k}_3) \rangle' = \left(2 \pi \right)^4P_\zeta^2  \frac{1}{(k_1k_2k_3)^2}S(k_1,k_2,k_3),
\ede
where $\zeta =-H\de\phi/\dot\phi_0$ is the curvature perturbation and $P_\zeta = H^4/(4\pi^2 \dot\phi_0^2)$ is the observed scalar power spectrum.
Then the dimensionless non-Gaussianity is given by
\begin{align}
S =\FR{6 f_\psi^2(X_0)H^2\dot\phi_0^2}{\pi^4}\bigg[C_\psi(\wt m)\Big(\FR{k_3}{2k_1}\Big)^{3-2\ii \wt m}+\text{c.c.}\bigg].
\end{align}
It can be easily checked that the presence of the Boltzmann suppression $e^{-2\pi\wt m}$ in $C_\psi(\wt m)$ when $\wt m\gg 1$. In the next section we shall see that the natural coupling for $f_\psi(X_0)$ increases with $\wt m$. Therefore, the non-Gaussianity calculated here is suppressed at small mass by small couplings while at large mass by the Boltzmann suppression, which means that this shape function can never be naturally large. To generate a large oscillatory signal in the squeezed bispectrum, we must also include the slow-roll correction, as we shall study in the next section.

\section{Neutrino Signatures on a Slow-Roll Background}
\label{sec_sl}

The analysis of the last section is basically in line with the methodology adopted in \cite{Chen:2016hrz}, where we only considered dS covariant corrections to the neutrino spectrum. However, as we shall elaborate in this section, it is the non-dS covariant slow-roll corrections to the neutrino spectrum that generate the most interesting effects, i.e. an amplified oscillatory signal in the squeezed limit as a result of a nonzero chemical potential generated by the inflaton background. We have already identified this boosted production in the mode function of neutrinos in Sec.\;\ref{sec_overview}. See discussions below (\ref{proplatetime}). In this section, we shall work out the corresponding neutrino spectrum in the context of EFT described by (\ref{OEFT}), in several different scenarios:
\begin{enumerate}
  \item General EFT (\ref{OEFT}); Electroweak symmetric phase.
  \item A reflective $\mathbb{Z}_2$ symmetry on the inflaton;  Electroweak symmetric phase.
  \item General EFT (\ref{OEFT}); Electroweak broken phase.
  \item Higgs portal couplings; Electroweak broken phase.
\end{enumerate}
This list is by no means exhaustive; we choose these scenarios simply because they are well-motivated general scenarios with distinct neutrino spectrum. With the method presented here, it would be straightforward to explore other possibilities in the future.

After examining the neutrino spectrum for the above list of scenarios, we shall then work out the 3-point function of scalar perturbations contributed by a neutrino loop. We shall not aim at a complete calculation of these loops as they can in general be extremely complicated. Instead, we shall adopt several simplifying approximations which works well in the squeezed limit and if we are only concerned with the oscillatory signals. As will be seen at the end of this section, the neutrino signals in the squeezed bispectrum can be made naturally large without fine tuning the parameters, even the signal itself is loop-suppressed.

\subsection{Slow-Roll Correction to Neutrino Spectrum}

In this subsection we shall assume that the electroweak symmetry is unbroken during inflation, and consider the neutrino spectrum after including all effective interactions in (\ref{OEFT}). In the symmetric phase, we have $\la h\ra=0$ and thus $\O_{hi}$ will be irrelevant in determining the neutrino mass, while some of $\O_{ni}$ will contribute to quadratic part of the neutrino Lagrangian which we collect as
\begin{align}
  \Delta\ld= &~\FR{\lam_{n1}\dot\phi_0}{\Lambda}\nu^\dag \ob\si^0 \nu+\FR{\lam_{n2}\dot\phi_0}{\Lambda}N^\dag \ob\si^0 N+\FR{\lam_{n6}H\dot\phi}{\Lambda^3}\nu^\dag\ii\ob{\sla{\D}}\nu+\FR{\lam_{n7}H\dot\phi}{\Lambda^3}N^\dag\ii\ob{\sla{\D}}N \n\\
  &~+\bigg(\FR{\lam_{n3}H\dot\phi_0}{\Lambda^2}+\FR{\lam_{n5}\dot\phi_0^2}{\Lambda^3}\bigg)\big(NN+\text{c.c.}\big).
\end{align}
After normalizing both fields, we see that the quadratic Lagrangian is in the same form as (\ref{quardldcp}) considered in Sec.\;\ref{sec_overview}, with the following mass parameters,
\begin{align}
  \lam_\nu=&-\bigg(1+\FR{\lam_{n6}H\dot\phi_0}{\Lambda^3}\bigg)^{-1}\FR{\lam_{n1}\dot\phi_0}{\Lambda},\\
  m_\nu =&~ 0,\\
  \lam_N=&-\bigg(1+\FR{\lam_{n7}H\dot\phi_0}{\Lambda^3}\bigg)^{-1}\FR{\lam_{n2}\dot\phi_0}{\Lambda},\\
  m_N =&~ \bigg(1+\FR{\lam_{n7}H\dot\phi_0}{\Lambda^3}\bigg)^{-1}\bigg(m_{N0} - \FR{\lam_{n3}H\dot\phi_0}{\Lambda^2}-\FR{\lam_{n5}\dot\phi_0^2}{\Lambda^3}\bigg).
\end{align}
Here $\lam_\nu$ and $\lam_N$ correspond to the $\lambda$-parameter in (\ref{quardldcp}) for $\nu$ and $N$, respectively, and so are $m_\nu$ and $m_N$. In addition, the left-handed neutrino will also receive a nonzero mass correction (\ref{Dmnu}) from the Higgs loop as described in the last section.

\paragraph{A special case with inflaton $\mathbb{Z}_2$ symmetry.}

When the inflaton is further subject to a reflective $\mathbb{Z}_2$ symmetry $\phi \to -\phi$, all $\mathcal{O}_{ni}$ in (\ref{OEFT}) but $\mathcal{O}_{n5}$ is forbidden. Therefore we are left with the following paramters,
\begin{align}
  &\lam_L=\lam_N =0, &&m_N=m_{N0}-\FR{\lam_{n5}\dot\phi_0^2}{\Lambda^3},
\end{align}
while $m_\nu$ is again given by (\ref{Dmnu}). Consequently, this scenario is identical to the situation considered in the last section, i.e., the correction to the neutrino spectrum, as well as the oscillatory signal, is dS covariant. This provides a simple proof of principle that the dS covariant scenario can be naturally realized in a general EFT setting, although the oscillatory signals in this scenario is unlikely to be large unless we tune the parameters.

\subsection{Neutrino Signatures with Electroweak Symmetry Breaking}

Next we consider a possible breaking of electroweak symmetry during inflation. The Higgs field during inflation may pick up a large VEV, comparable or much greater than the Hubble scale. For example, the unique dimension-4 non-minimal coupling between the Higgs field and the Ricci scalar can generate a Higgs VEV $h_0\sim \order{H}$ since the Ricci scalar $R=12H^2$ during inflation. As a second example, a negative effective operator $\mathcal{O}_{h2}$ in (\ref{Oh}) would also generate a nonzero Higgs VEV, $h_0 \sim \dot\phi_0/(\sqrt{\lam}\Lambda)$ where $\lam$ is the quartic self-coupling strength of the Higgs field.

In the case of nonzero Higgs VEV, all operators $\mathcal{O}_{ni}$ in (\ref{On}) contribute to the quadratic Lagrangian of the neutrinos. After evaluating on the nonzero inflaton and Higgs background, the quadratic Lagrangian becomes
\begin{align}
  \ld= &~\bigg(1 + \FR{\lam_{n6}H\dot\phi}{\Lambda^3}\bigg)\nu^\dag\ii\ob{\sla{\D}}\nu+a\bigg(\FR{\lam_{n1}\dot\phi_0}{\Lambda}+\FR{\lam_{n10}h_0^2\dot\phi_0}{\Lambda^3}\bigg)\nu^\dag \ob\si^0 \nu\n\\
  &~+\bigg(1+\FR{\lam_{n7}H\dot\phi}{\Lambda^3}\bigg)N^\dag\ii\ob{\sla{\D}}N+a\bigg(\FR{\lam_{n2}\dot\phi_0}{\Lambda}+\FR{\lam_{n11}h_0^2\dot\phi_0}{\Lambda^3}\bigg)N^\dag \ob\si^0 N \n\\
  &~-\FR{1}{2}a\bigg(m_{N0}-\FR{2\lam_{n3}H\dot\phi_0}{\Lambda^2}-\FR{2\lam_{n5}\dot\phi_0^2}{\Lambda^3}\bigg)\big(NN+\text{c.c.}\big)\n\\
  &~-a\bigg(y_n h_0-\FR{\lam_{n8}Hh_0\dot\phi_0}{\Lambda^3}\bigg)(\nu N+\text{h.c.})+a\FR{\lam_{n4}h_0\dot\phi_0}{\Lambda^2}(\nu^\dag\bar\si^0 N+\text{h.c.})\n\\
  &~-\FR{\lam_{n9}h_0\dot\phi_0}{\Lambda^3}(\nu N'+\text{h.c.}).
\end{align}
We will not consider the most general case, but only consider the cases where the mass correction is not too greater than the Hubble scale, which means that the cutoff scale is high and that we can retain dim-5 operators only. Most of above terms disappear then, leaving the following simplified Lagrangian,
\begin{align}
  \ld= &~ \nu^\dag\ii\ob{\sla{\D}}\nu+a \FR{\lam_{n1}\dot\phi_0}{\Lambda} \nu^\dag \ob\si^0 + N^\dag\ii\ob{\sla{\D}}N+a \FR{\lam_{n2}\dot\phi_0}{\Lambda} N^\dag \ob\si^0 N -\FR{1}{2}a m_{N0} \big(NN+\text{c.c.}\big)\n\\
  &~-a y_n h_0 (\nu N+\text{h.c.}).
\end{align}
We will further assume $\lam_{n1}=\lam_{n2}$ for simplicity. Then the mass matrix can be diagonalized as usual, and the Lagrangian can be rewritten in terms of mass eigenstates, which we call $N_\pm$,
\begin{align}
  \ld= &~ N_+^\dag\ii\ob{\sla{\D}}N_+ + N_-^\dag\ii\ob{\sla{\D}}N_- -a \lam \big(N_+^\dag \ob\si^0 N_+ + N_-^\dag \ob\si^0 N_-\big) -\FR{1}{2} \big(m_+ N_+ N_+ + m_- N_- N_- +\text{c.c.}\big),
\end{align}
where
\begin{align}
\label{nspecEWSB}
  & \lam = \FR{\lam_{n1}\dot\phi_0}{\Lambda},
  &&m_\pm = \FR{m_{N0}\pm\sqrt{m_{N0}^2+4h_0^2}}{2}.
\end{align}
When $m_{N0}\gg h_0$, the lighter mass eigenvalue becomes $m_-\simeq -h_0^2/m_{N0}$ which recovers the well-known seesaw mechanism.

\paragraph{Neutrino Signatures with Higgs-Portal Coupling.}

Finally there is another option that the SM communicates with the inflaton field mostly through a Higgs portal coupling. Since the inflaton has to interact with the Higgs through derivative couplings as required by the approximate shift symmetry, the operator for the Higgs-inflaton interaction, unlike the more familiar dark matter Higgs-portal coupling, must have a dimension higher than 4, and thus must be associated with a scale. Then, the inflaton Higgs-portal coupling can be realized in a scenario where the theory describing the Higgs-inflaton sector has a much lower cutoff scale than the effective couplings between the inflaton and other SM fields. Then, the spectrum of SM fields, apart from the Higgs field, is much less affected by the slow-roll background of the inflaton. There is of course a question of how the separation of two scales can be naturally maintained. But here we are not going to explore the model building aspects of this scenario any further, but focus on its consequences on the neutrino signatures.

To mediate the signatures of fermions and vector bosons in SM to inflaton correlations at tree level, it is required that the Higgs develops a nonzero VEV. Consequently, the neutrino masses will be identical to (\ref{nspecEWSB}), while the chemical potential $\lam$ will be zero. The neutrino signals in the squeezed bispectrum in this case will be quite different from other cases since the neutrino communicates with the inflaton only through the Higgs. We shall consider this process later in this section.

\subsection{Signatures in the Bispectrum}

Now we study the neutrino signatures in the squeezed bispectrum, taking account of a nonzero chemical potential $\lam$. the result here applies to both the electroweak symmetric phase and the broken phase. Again, the leading contribution comes from 1-loop diagrams, while the couplings between the inflaton perturbation $\de\phi$ and the neutrinos are now from the dim-5 operators $\mathcal{O}_{n1}$ or $\mathcal{O}_{n2}$ in (\ref{OEFT}). Therefore the corresponding 1-loop diagram has three internal legs, as shown below.
\bge
\label{3ptloop}
  \parbox{0.35\textwidth}{\includegraphics[width=0.35\textwidth]{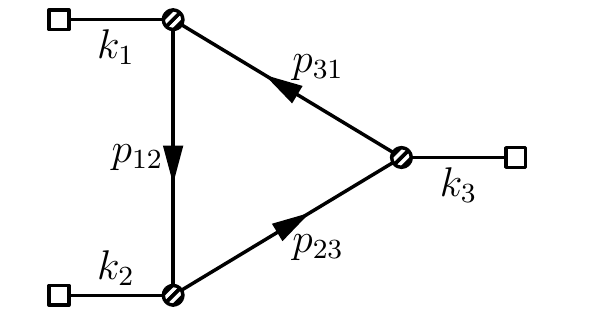}}~~+~~
  \parbox{0.35\textwidth}{\includegraphics[width=0.35\textwidth]{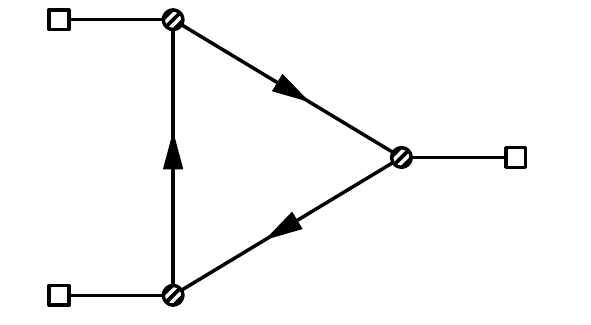}}
\ede
Here we have adopted the diagrammatic representation reviewed in \cite{Chen:2017ryl}. The external lines terminating in squares denote the inflaton field, while the internal line with a single arrow represents the neutrino propagator $D_{ab\al\dot\be}$. In addition, we note that each vertex decorated with a shaded circle represents two possible choices of in or out type in the diagrammatic approach of in-in formalism.  More details about neutrino propagators in the in-in formalism are summarized in App.\;\ref{app_sk}.

Using the diagrammatic rule of the in-in formalism (see  \cite{Chen:2017ryl} for a review), we have, for the sum of two diagrams in (\ref{3ptloop}),
\begin{align}
\label{3ptint}
&\la\de\phi(\mb k_1)\de\phi(\mb k_2)\de\phi(\mb k_3)\ra'_{\mathcal{O}_{n2}} \n\\
=&  \sum_{a,b,c=\pm} abc\Big(\FR{\ii}{\Lambda}\Big)^3\int_{-\infty}^0\di\tau_1\di\tau_2\di\tau_3\,\mathcal{F}_{\mu a}(\mb k_1,\tau_1)\mathcal{F}_{\nu  b}(\mb k_2,\tau_2)\mathcal{F}_{\lam c}(\mb k_3,\tau_3)\int\FR{\di^3\mb q}{(2\pi)^3}\, \mathcal{T}^{\mu\nu\lam}_{abc},
\end{align}
where the sum over $a,b,c$ takes account of all in-in contours, the factor $(\ii/\Lambda)^3$ comes from the couplings of the three vertices. Here we have put $\lambda_{n1}=\lambda_{n2}=1$ in (\ref{OEFT}) without loss of generality, since we can always rescale the cutoff $\Lambda$ to make the dimensionless couplings equal to 1. (We remind the reader that $\lambda_{n1}=\lambda_{n2}$ is assumed a priori for simplicity, as we did above.) The vector function $\mathcal{F}_{\mu a}(\mb k,\tau)$ comes from the external leg of the inflaton field $\de\phi(\mb k)$, and is given by
\bge
\mathcal{F}_{\mu a}(\mb k,\tau)\equiv \bgp \pd_{\tau_1} G_a(k;\tau) \\ \ii\mb k G_a(k;\tau)\edp  =  \FR{H^2}{2k^3}\bgp k^2\tau \\ \ii\mb k(1-\ii a k\tau)\edp e^{+\ii a k\tau},
\ede
in which $G_a(k;\tau)$ denote the boundary-to-bulk propagator of the inflaton field \cite{Chen:2017ryl},
\bge
  G_\pm (k;\tau)= \FR{H^2}{2k^3}(1\mp \ii k \tau)e^{\pm \ii k\tau},
\ede
and the trace $\mathcal{T}_{abc}^{\mu\nu\lam}$ comes from the spinor loop, and is given by
\begin{align}
  \mathcal{T}_{abc}^{\mu\nu\lam}
  =&  -\tr\Big[\bar\si^{\mu\dot\al\al}D_{ab \al\dot\be}(\mb p_{12},\tau_1,\tau_2)\bar\si^{\nu\dot\be\be}D_{bc \be\dot\ga}(\mb p_{23},\tau_2,\tau_3)\bar\si^{\lam\dot\ga\ga}D_{ca \ga\dot\al}(\mb p_{31},\tau_3,\tau_1)\Big]\n\\
  & -\tr\Big[\bar\si^{\mu\dot\al\al}D_{ac \al\dot\ga}(\mb p_{13},\tau_1,\tau_3)\bar\si^{\lam\dot\ga\ga}D_{cb \ga\dot\be}(\mb p_{32},\tau_3,\tau_2)\bar\si^{\nu\dot\be\be}D_{ba \be\dot\al}(\mb p_{21},\tau_2,\tau_1)\Big],
\end{align}
with $\mb p_{12}=-\mb p_{21}=\mb q$, $\mb p_{23}=-\mb p_{32}=\mb q+\mb k_2$, $\mb p_{31}=-\mb p_{13}=\mb q-\mb k_1$, and $\mb q$ is the loop momentum. We will simplify the calculation by choosing a special configuration for the external momenta,
\begin{align}
  &-\mb k_2 \simeq \mb k_1 = (0,0,k_1 ),
  &&\mb k_3=(0,0,k_3).
\end{align}
There may also be nontrivial angular dependence if we allow an arbitrary angle between $\mb k_1$ and $\mb k_3$, which we will consider elsewhere in the future.

In general, it is very difficult to carry out all integrals completely  in (\ref{3ptint}). But several simplifying assumptions can be made if we are only interested in the oscillatory signals in the squeezed bispectrum.

Firstly, we shall assume that we are working in the parameter region  $m\sim$ or $<H$, and $\lam \gg H$. In this region the positive helicity states dominate the propagator in the soft limit $|k\tau|\ll 1$ and the $h_-h_-^\dag$ component of the propagator is suppressed by the Boltzmann-like factor $e^{-2\pi\wt\lam}$ which can thus be neglected. Alternatively, were we take negative chemical potential $\lam<0$, the positive helicity states should be neglected instead.

Secondly, we shall assume the squeezed limit $k_3\ll k_{1,2}$ and we shall focus on the clock signal only. This allows us to approximate the loop integral by its value near a configuration where two of three internal legs become very soft, with momentum $\sim k_3/2$, and the other leg remains hard, with momentum $\sim k_1$. Therefore, we shall expand the two internal lines in late-time limit $|k_1\tau|\ll 1$. Meanwhile, since the clock signal is mostly generated when the resonant time-integrand is on its saddle point, which means that $|2k_1\tau|\simeq m/H$, we shall just evaluate the hard internal line at $|2k_1\tau|=m/H$. Therefore we will not do a complete loop integral, but only restrict ourselves to this soft configuration. Equivalently, one may think of our computation as introducing a loop-momentum cutoff $\Lambda_q\simeq k_3$ such that $|\mb p_{23}|,|\mb p_{31}|\leq \Lambda_q$. As a result, we will simply approximate the loop integral measure by
\bge
  \int\FR{\di^3\mb q}{(2\pi)^3} \simeq \FR{k_3^3}{(2\pi)^3}\int_0^{2\pi}\di\varphi,
\ede
where we have kept the azimuthal integral as it is not difficult to leave the azimuthal angle $\varphi$ free in our choice of loop momentum configuration. Given the condition that $k_{1,2}\gg k_3$, and that $p_{23}\simeq p_{31}\simeq k_3$, we see that $\mb p_{12}\simeq \mb k_1$. Thus a convenient choice of loop momentum configuration is the following
\begin{align}
  &\mb p_{12}\simeq \mb k_1,
  &&\mb p_{23}\simeq k_3(\fr{\sqrt{3}}{2}\cos\varphi, \fr{\sqrt 3}{2}\sin\varphi,-\fr{1}{2}),
  &&\mb p_{31}\simeq k_3(\fr{\sqrt{3}}{2}\cos\varphi, \fr{\sqrt 3}{2}\sin\varphi,\fr{1}{2}).
\end{align}
To proceed, we make use of the late-time limit of the fermion propagator for the two soft lines (those carry $\mb p_{23}$ and $\mb p_{31}$), taken from $h_+^{}h_+^\dag$ component of (\ref{proplatetime}),
\begin{align}
  D_{ab \al\dot\be}(\mb k;\tau_1,\tau_2)
  =&~2\,\text{Re}\,\bigg[\Big(f_1h_+^{}(\mb k)h_+ ^\dag(\mb k)
  +f_2h_-^{}(\mb k)h_+ ^\dag(\mb k) \Big)(4k^2\tau_1\tau_2)^{+\ii\wt\mu} \bigg], \\
  f_1\equiv&~\FR{-e^{\pi\wt\lam}\Gamma^2(-2\ii\wt\mu)}{\Gamma(\ii\wt\lam-\ii\wt\mu)\Gamma(-\ii\wt\lam-\ii\wt\mu)},\\
  f_2\equiv&~\FR{\wt m\Gamma^2(-2\ii\wt \mu)}{\Gamma(1-\ii\wt\lam-\ii\wt\mu)\Gamma(-\ii\wt\lam-\ii\wt\mu)},
\end{align}
For the hard internal line ($\mb p_{12}$), we evaluate the propagator at $|2k\tau|=\wt m$. When $\wt\lam\gg\wt\mu$, the result is
\begin{align}
   &~D_{ab \al\dot\be} \simeq g_{++} h_+(\mb k) h_+^\dag (\mb k)+g_{+-} h_+(\mb k) h_-^\dag (\mb k)+g_{-+} h_-(\mb k) h_+^\dag (\mb k)+g_{--} h_-(\mb k) h_-^\dag (\mb k), \\
   \label{gpp}
   &~g_{++}= \wt m^2\bigg[\,\text{Ci}^2(\wt m)+\Big(\FR{\pi}{2}-\text{Si}(\wt m)\Big)^2\bigg],\\
   \label{gpm}
   &~g_{+-}= g_{-+}^* = -\wt m e^{-\ii\wt m}\Big[\text{Ei}(\ii\wt m)-\ii\pi\Big],\\
   \label{gmm}
   &~g_{--}=1,
\end{align}
where $\wt m=m/H$ as in (\ref{lammu}), Si$(z)=\int_0^z\,t^{-1}\sin t\di t$ is the sine integral function, Ci$(z)=\int_{-z}^\infty t^{-1}\cos t\di t$ is the cosine integral function, and Ei$(z)=-\int_{-z}^\infty t^{-1} e^{-t}\di t$ is the exponential integral function. The $\pm$ indices in $g$-factors above indicate the helicity components and should not be confused with the in-in indices. The helicity eigenspinor corresponding to a momentum in the direction of $(\sin\theta\cos\varphi,\sin\theta\sin\varphi,\cos\theta)$ can be chosen as
\begin{align}
   &h_+(\theta,\varphi)=\bgp \cos\frac{\theta}{2} \\[2mm] e^{\ii\varphi}\sin\frac{\theta}{2} \edp,
   &&h_-(\theta,\varphi)=\bgp -e^{-\ii\varphi}\sin\frac{\theta}{2} \\[2mm] \cos\frac{\theta}{2} \edp.
\end{align}
With all these ingredients known, now we can carry out all integrals in (\ref{3ptint}). The result is,
\begin{align}
  \la\de\phi^3\ra'_{\mathcal{O}_{n2}}
  =&~ 2\text{Re}\,\bigg[\FR{1}{(2\pi)^3} \Big(\FR{\ii}{\Lambda}\Big)^3\FR{3H^6}{16\pi^2k_1^6}\bigg(\FR{4k_3}{k_1}\bigg)^{2\ii\wt\mu}\bigg(\sum_{a,b=\pm}g_{ab}f_{ab}\bigg)\bigg],
\end{align}
where $g_{ab}$ are given in (\ref{gpp})-(\ref{gmm}), and $f_{ab}$ are given by
\begin{align}
  f_{++}=f_{--}=&~\wt \mu e^{2\pi\wt\lam}\sinh^2[\pi(\wt\lam-\wt\mu)]\sinh^2[\pi(\wt\lam+\wt\mu)]\n\\
  &~\times\Gamma^2(1-\ii\wt\lam+\ii\wt\mu)\Gamma^3(1-2\ii\wt\mu)\Gamma^2(\ii\wt\mu)\Gamma^2(1+\ii\wt\lam+\ii\wt\mu),\\
  f_{+-}=&~ \wt\mu e^{2\pi\wt\lam}\FR{\pi^4\Gamma^3(1-2\ii\wt\mu)\Gamma^2(\ii\wt\mu)}{2\Gamma^2(\ii\wt\lam-\ii\wt\mu)\Gamma^2(-\ii\wt\lam-\ii\wt\mu)},\\
  f_{-+}=&-\FR{32\ii\pi}{\Gamma(1-\ii\wt\lam-\ii\wt\mu)}\wt m\wt\mu^5e^{\pi \wt\lam }\sinh[\pi(\wt\lam-\wt\mu)]\sinh^2[\pi(\wt\lam+\wt\mu)]\n\\
  &~\times\Gamma(1-\ii\wt\lam+\ii\wt\mu)\Gamma^2(\ii\wt\mu)\Gamma^3(-2\ii\wt\mu)\Gamma^2(1+\ii\wt\lam+\ii\wt\mu).
\end{align}
We can rewrite the result, using the dimensionless shape function $S$, defined in (\ref{3ptshaperaw}), as
\begin{align}
\label{NloopShape}
 &  S=-\FR{3}{32\pi^3}\wt\lam^3P_\zeta\times 2\text{Im\,}\bigg[\wt C_\psi\bigg(\FR{k_3}{k_1}\bigg)^{2+2\ii\wt\mu}\bigg],\\
 & \wt C_\psi\equiv 2^{4\ii\mu}\sum_{a,b=\pm}g_{ab}f_{ab},
\end{align}
where we have replaced the cutoff scale $\Lambda$ by $\wt\lam$ via $\wt\lam=\lam/H=\dot\phi_0/(H\Lambda)$.

We note that the above result is valid only when the dimensionless chemical potential $\wt\lam \gg 1$. Given $\wt\lam = \dot\phi / (H\Lambda)$, this corresponds to $\Lambda < \dot\phi/H$. We note that the unitarity put a lower bound on $\Lambda$ roughly as $\Lambda > \dot\phi_0^{1/2}$, which means that $\wt\lam$ can be as large as $\dot\phi_0^{1/2}/H$. Thus we see that there is a range of viable parameter space for the above expression to be valid. To see the magnitude of this clock signal, we use the expansion $\Gamma(x+\ii y)\sim \sqrt{2\pi }|y|^{x-1/2}e^{-\pi |y|/2}$ with $x\geq 0$ and $y$ real, which is valid when $y\gg 1$ and holds up to a pure phase, and also $g_{\wt m}\simeq 1$. Then, the non-Gaussianity $f_{NL}$, which we define to be the absolute value of the coefficient of $(k_3/k_1)^{2+2\ii\wt\mu}$, can be approximated by
\bge
\label{fNLapp}
  f_{NL}(\text{clock})\simeq \FR{3\pi^2}{2}P_\zeta \wt\lam^5\wt m^3 e^{-5\pi\wt m^2/(4\wt\lam)}.
\ede
From this expression we see clearly that the non-Gaussianity is enhanced by powers of $\wt\lam$ for large $\wt\lam$ rather than suppressed exponentially. Therefore even we have a great suppression from the factor $P_\zeta\simeq 2\times 10^{-9}$ in $f_{NL}$, the result can still be greatly enhanced by choosing a large $\wt\lam$. 

\begin{figure}[t]
\centering
\includegraphics[width=0.55\textwidth]{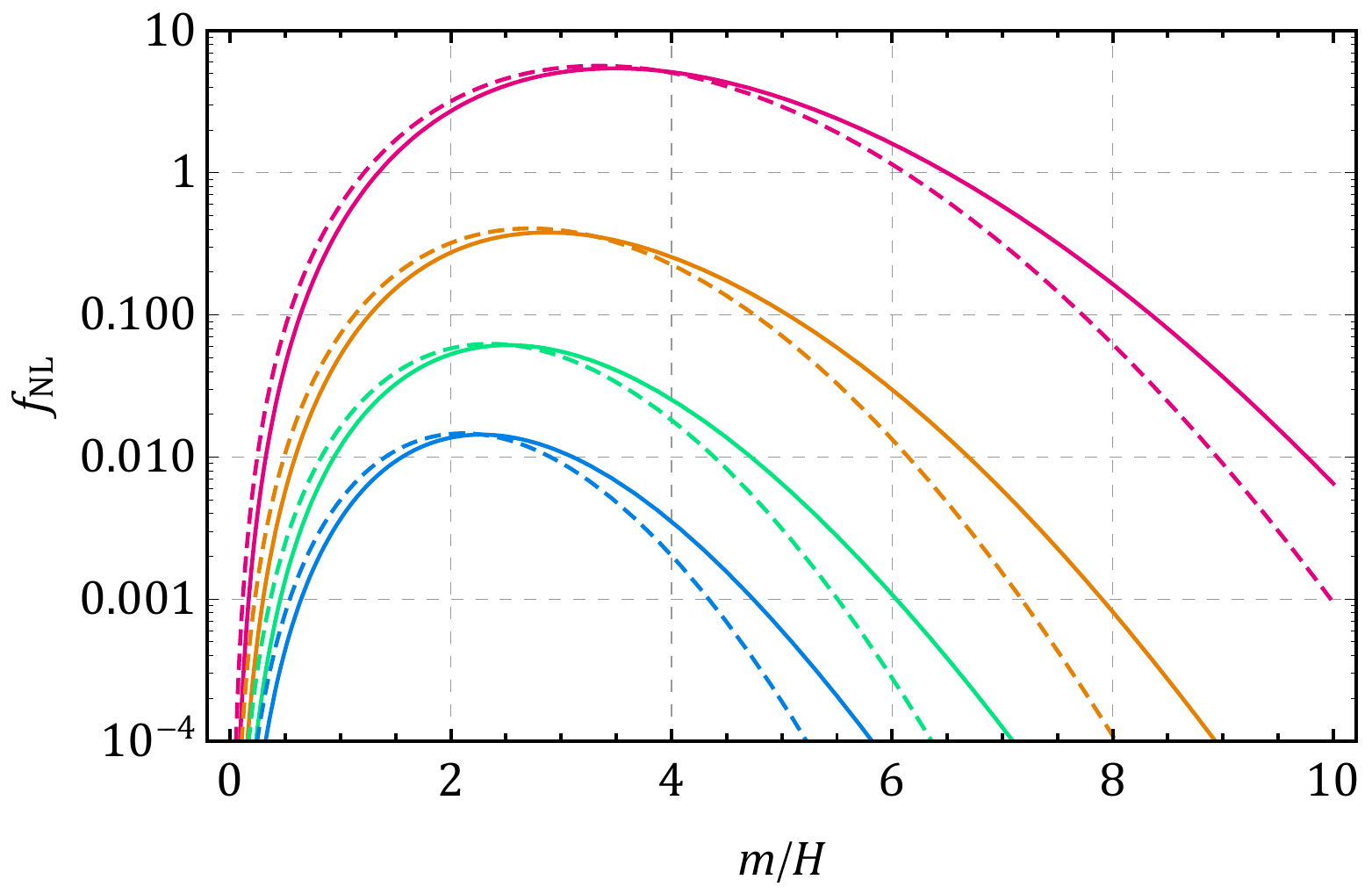}
\caption{The dimensionless strength of the clock signals, as functions of the fermion mass $\wt m=m/H$. The 4 solid curves from bottom to top correspond to choosing $\Lambda = (5,4,3,2)\dot\phi_0^{1/2}$, respectively, and the 4 dashed curves are corresponding approximate results of (\ref{fNLapp}).}
\label{FigfNL}
\end{figure}

\begin{figure}[t]
\centering
\includegraphics[width=0.8\textwidth]{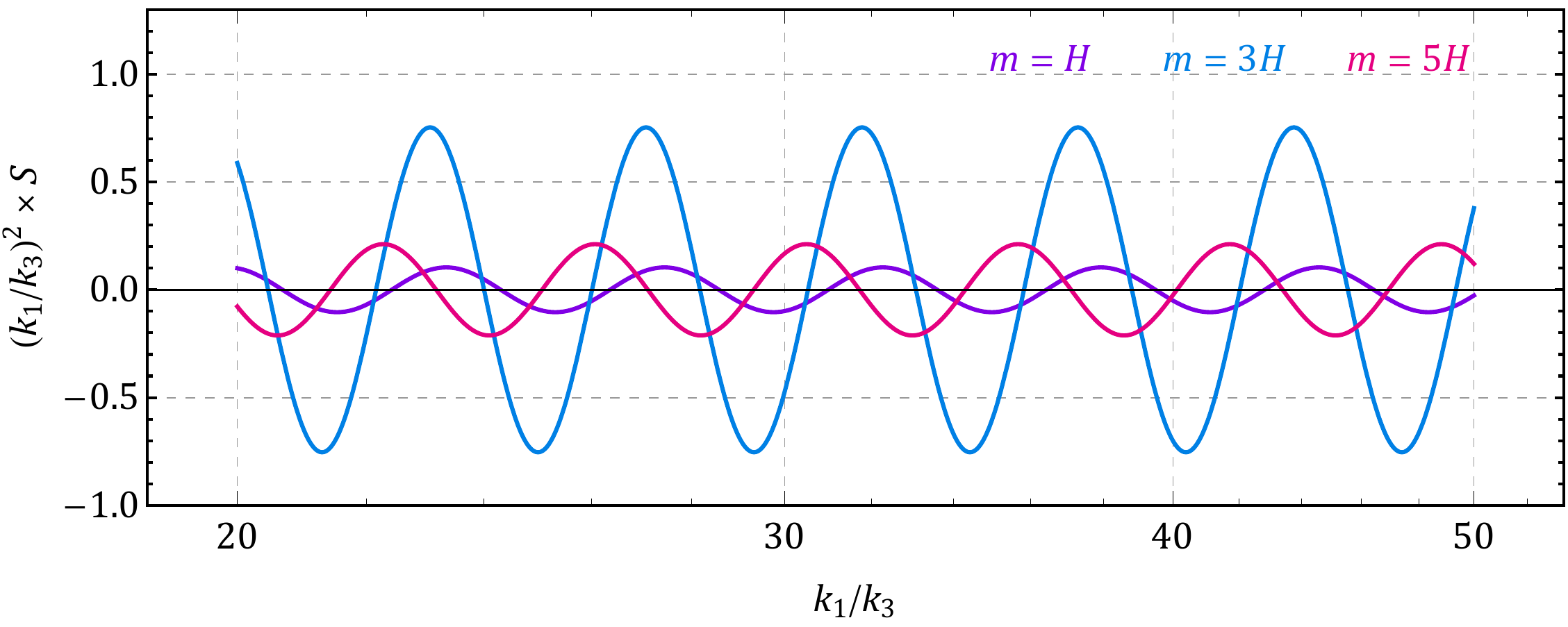}
\caption{The clock signals $S$ in (\ref{NloopShape}) as functions of the momentum ratio $k_1/k_3$. In this plot $\Lambda=3\dot\phi_0^{1/2}$ and the blue, orange, and green curves correspond to $m = (1,3,5)H$, respectively.}
\label{Fig_S}
\end{figure}

In Fig.\;\ref{FigfNL} we plot $f_{NL}$ from (\ref{NloopShape}) and the approximation (\ref{fNLapp}) for several choices of $\Lambda$ (or equivalently, for $\wt\lam$) and for a range of mass $m$.  In addition, we plot directly the oscillatory signals for $\Lambda=\dot\phi_0^{1/2}$ and for $m=(1,3,5)H$ in Fig.\;\ref{Fig_S}. It can be seen that the non-Gaussianity of the neutrino signal can be made close to $\order{1}$ once the cutoff scale is lowered towards its unitarity bound.  We stress again that there is no fine tuning of parameters in this scenario and thus the clock signal is made observably large in a technically natural way.

The above calculation considered only the clock signals, namely the oscillatory part of the bispectrum. We showed that the non-Gaussianity $f_{NL}$ of the clock signal can be made large within a natural EFT. Given a large clock signal $\sim\order{1}$, one may worry that the overall $f_{NL}$, namely the amplitude of the non-clock part of the bispectrum, would be too large to be consistent with the current Planck constraint \cite{Ade:2015ava}. However, the overall $f_{NL}$ can be roughly estimated to be $\sim P_\zeta \wt\lam^3/(4\pi)^2$ for $\wt m\gtrsim\order{1}$. Comparing this with the clock signal (\ref{fNLapp}), we see that the non-clock part does not dominate for a wide range of parameters we are interested in, and a large clock signal is not in conflict with the Planck constraint.

\paragraph{Neutrino Signatures from Higgs-Portal Coupling. }

Now we consider the neutrino signals in the case of Higgs portal coupling and we estimate the strength of the non-Gaussianity from the neutrino loop in the squeezed bispectrum, although there remains a question of how to realize the Higgs-portal coupling scenario in a technically natural way at the first place.

The 1-loop process involving neutrinos in this scenario is similar to the one considered before, except that the neutrino couples to Higgs rather than to the inflaton perturbation. Instead, the Higgs is mixed with the inflaton from a bilinear insertion, as shown below.
\bge
\parbox{0.4\textwidth}{\includegraphics[width=0.4\textwidth]{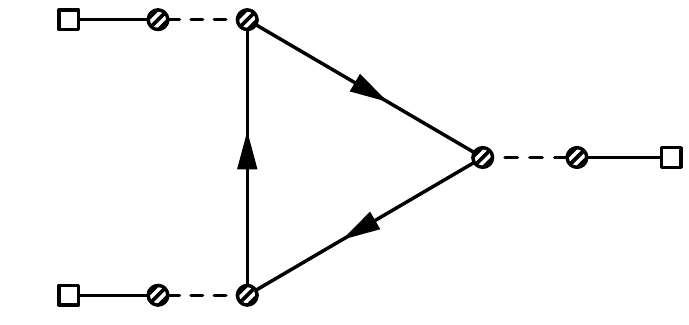}}
\ede
We shall not spell out all the technical details of computing this diagram, which is largely the same with the previous one. For our purpose, it suffices to estimate the magnitude of the corresponding non-Gaussianity $f_{NL}$ as
\begin{align}
  f_{NL}(\text{clock})\simeq \FR{1}{16\pi^2}\FR{1}{P_\zeta^{1/2}}\bigg(\FR{h_0}{\Lambda}\bigg)^3\bigg(\FR{H}{m_h}\bigg)^3\bigg(\FR{\sqrt 2 m }{h_0}\bigg)^3\FR{\Gamma^4(-2\ii\wt m)}{\Gamma^4(-\ii\wt m)},
\end{align}
in which $1/(16\pi^2)$ is the loop factor; $1/P_\zeta^{1/2}$ comes from the definition of the shape function; $(h_0/\Lambda)^3$ comes from three bilinear mixings between the inflaton and the Higgs, derived from the dimension-5 operator $(\pd_\mu\phi)\mb H\pd^\mu \mb H/\Lambda$; $(H/m_h)^3$ comes from the Higgs propagator and it is assumed that $m_h\geq H$; $(\sqrt{2}m/h_0)^3$ are three Yukawa couplings; and finally, the $\Gamma$-function factors are from the late-time behavior of two soft neutrino propagators. When $m\sim m_h\sim h_0\sim\order{H}$, which can be naturally realized since the dimension-5 operator mentioned above does not contribute to Higgs VEV, we see that the above expression is suppressed by the loop factor and by the cutoff scale $(H/\Lambda)^3\gtrsim (H/\dot\phi_0^{1/2})^3$. Therefore the non-Gaussianity in this case is expected to be small and is $\sim P_\zeta$.

\section{Discussions}
\label{sec_concl}

In this paper we have studied systematically the inflationary mass spectrum of neutrinos, both left-handed and right-handed, and also their characteristic signatures in the squeezed bispectrum. We considered a single family of neutrinos in the type I seesaw model where the right-handed neutrino can be a priori heavy. We couple this model with the inflaton in a generic single-field inflation in an EFT manner.

The most important result of the paper is that the slow-roll background of the inflaton can provide a large chemical potential for the neutrino at the leading order effective inflaton-neutrino coupling, which can greatly amplify the oscillatory signal of the neutrino in the squeezed bispectrum. The corresponding non-Gaussianity $f_{NL}$ can be of $\order{1}$ in a general EFT without fine tuning even it appears from at least one loop. The rolling inflaton as a chemical potential may also be realized for other types of fields, scalars and vector bosons, and thus is quite significant in the cosmological collider physics. We leave a more general study of this problem in the future.

The result of this paper can be easily generalized to any massive singlet fermions, or ``sterile neutrinos'', that couple to the inflaton field. As we have seen in this paper, the tree-level mass of the fermion is crucial in generating the oscillatory signal: the signal would vanish if the tree-level (Majorana) mass is sent to zero. Therefore, if the electroweak symmetry is not broken during inflation, the neutrinos are the only candidate in SM that can generate such a signal. On the other hand, it is an interesting open question that how to know the singlet fermion with a Majorana mass is indeed the right-handed neutrino taking part in the seesaw mechanism. We leave this question for future studies.

\paragraph{Acknowledgements.} We thank Xunjie Xu for initial collaboration on this project. YW thanks Henry Tye and Tao Liu for discussions.
XC is supported in part by the NSF grant PHY-1417421. YW is supported in part by ECS Grant 26300316 and GRF Grant 16301917 from the Research Grants Council of Hong Kong. ZZX is supported in part by Center of Mathematical Sciences and Applications, Harvard University.

\begin{appendix}

\section{Weyl Spinors in de Sitter Spacetime}
\label{app_dS}


In this appendix we solve the equations for the neutrino propagators (\ref{GreenEqn1}) and (\ref{GreenEqn2}), with the parameterization given in (\ref{spinorprop1}) and (\ref{spinorprop2}).  To this end, we need to know how to take derivatives of the spinor translator $P_\al{}^{\al'}$. Making use of the symmetry, we can parameterize its derivative as
\bge
\label{DD}
\si^\mu_{\be\dot \be}\D_\mu P_\al{}^{\al'}=u(L)\si^\mu_{\be\dot \be}n_\mu P_\al{}^{\al'}+v(L)\si^\mu_{\al\dot \be}n_\mu P_\be{}^{\al'},
\ede
where $L$ is the geodesic distance between $x$ and $x'$. Then the problem boils down to finding scalar functions $u(L)$ and $v(L)$. We start from the definition $n^\mu\D_\mu P_\al{}^{\al'}=0$, and rewrite it in the following form using the relation $\bar\si^{\mu\dot \be\be}\si^\nu_{\be\dot \be}=-2g^{\mu\nu}$,
\bge
  \bar\si^{\mu\dot\be\be}\si^\nu_{\be\dot\be}n_\mu\D_\nu P_\al{}^{\al'}=0.
\ede
Then the left hand side becomes
\bge
  -2g^{\mu\nu}n_\mu\D_\nu P_\al{}^{\al'}=-2u(L) P_\al{}^{\al'}-v(L)\de^\be{}_\al P_\be{}^{\al'}.
\ede
Therefore we see that $u(L)=-\frac{1}{2}v(L)$ and
\bge
  \si^\mu_{\be\dot \be}\D_\mu P_\al{}^{\al'}=v(L)\Big(-\FR{1}{2}\si^\mu_{\be\dot \be}n_\mu P_\al{}^{\al'}+\si^\mu_{\al\dot \be}n_\mu P_\be{}^{\al'}\Big).
\ede
To determine $v(L)$, we use the property of covariant derivatives
\bge
  [\D_\mu,\D_\nu]\phi_A(x)=\FR{\ii}{2}R_{\mu\nu}{}^{\rh\si}(S_{\rh\si})_A{}^{B}\phi_B(x),
\ede
where $\phi_A(x)$ is any field charged in a linear representation of Lorentz group, and $(S_{\rh\si})_A{}^B$ is the corresponding representation matrix of Lorentz generator. For left handed spinor, we have
\bge
  (S^{\mu\nu})_\al{}^\be=\FR{\ii}{4}(\si^\mu\bar\si^\nu-\si^\nu\bar\si^\mu)_\al{}^\be.
\ede
Therefore we have
\bge
\label{app1com}
  [\D_\mu,\D_\nu]P_\al{}^{\al'}=-\FR{1}{4}H^2(\si_\mu\bar\si_\nu-\si_\nu\bar\si_\mu)_\al{}^{\be}P_\be{}^{\al'}.
\ede
So now we evaluate the following expression
\begin{align}
  &\si^\mu_{\al\dot\al}\si^\nu_{\be\dot\be}\D_\mu\D_\nu P_\ga{}^{\ga'}
  =\si^\mu_{\al\dot\al}n_\mu v'(L)\big(-\fr{1}{2}\si^\nu_{\be\dot\be}n_\nu P_\ga{}^{\ga'}+\si^\nu_{\ga\dot\be}n_\nu P_\be{}^{\ga'}\big)\n\\
   &~+v(L)\big[-\fr{1}{2}\si^\mu_{\al\dot\al}\si^\nu_{\be\dot\be}A(g_{\mu\nu}-n_\mu n_\nu)P_\ga{}^{\ga'}-\fr{1}{2}\si^\nu_{\be\dot\be} n_\nu v(L)\big(-\fr{1}{2}\si_{\al\dot\al}^\mu n_\mu P_\ga{}^{\ga'}+\si^\mu_{\ga\dot \al}n_\mu P_\al{}^{\ga'}\big)\n\\
   &~+\si^\mu_{\al\dot \al}\si^\nu_{\ga\dot\be}A(g_{\mu\nu}-n_\mu n_\nu)P_\be{}^{\ga'}+\si^\nu_{\ga\dot\be}n_\nu v(L)\big(-\fr{1}{2}\si^\mu_{\al\dot \al}n_\mu P_\be{}^{\ga'}+\si^\mu_{\be\dot \al}n_\mu P_\al{}^{\ga'}\big)\big],
\end{align}
from which we have
\begin{align}
\label{app1left}
\si^\mu_{\al\dot\al}\si^\nu_{\be\dot\be}[\D_\mu,\D_\nu] P_\ga{}^{\ga'}
=&~(\si^\mu_{\al\dot\al}\si^\nu_{\be\dot\be}-\si^\mu_{\be\dot\be}\si^\nu_{\al\dot \al})\D_\mu\D_\nu P_\ga{}^{\ga'}\n\\
=&~\FR{1}{2}(v'+3Av+v^2)\big(\ep_{\ga\be}\ep_{\dot \al\dot\be}P_\al{}^{\ga'}-\ep_{\al\ga}\ep_{\dot\al\dot\be}P_\be{}^{\ga'}\big).
\end{align}
On the other hand,
\bge
\label{app1right}
 -\FR{1}{4}H^2\si^\mu_{\al\dot\al}\si^\nu_{\be\dot\be}(\si_\mu\bar\si_\nu-\si_\nu\bar\si_\mu)_\ga{}^{\de}P_\de{}^{\ga'}
 =H^2\big(\ep_{\ga\be}\ep_{\dot\al\dot\be}P_\al{}^{\ga'}-\ep_{\al\ga}\ep_{\dot\al\dot\be}P_\be{}^{\ga'}\big).
\ede
Equating (\ref{app1left}) and (\ref{app1right}) according to (\ref{app1com}), we get
\bge
  v'+3Av+v^2-2H^2=0,
\ede
from which we can solve $v$ to be
\bge
  v(L)=-A-C=H\tan(HL/2),
\ede
given the condition $v(L=0)=0$. Therefore,
\bge
\si^\mu_{\be\dot\be}\D_\mu P_\al{}^{\al'}=H\tan(HL/2)\Big(-\FR{1}{2}\si^\mu_{\be\dot \be}n_\mu P_\al{}^{\al'}+\si^\mu_{\al\dot\be}n_\mu P_\be{}^{\al'}\Big).
\ede

Then the first equation (\ref{GreenEqn1}) gives
\bge
  f'(L)+\FR{3}{2}(A-C)f(L)=\ii m g(L).
\ede
The second equation (\ref{GreenEqn2}) gives
\bge
  g'(L)+\FR{3}{2}(A+C)g(L)=-\ii m f(L).
\ede
Eliminating $g$, we get
\bge
  f''(L)+3Af'(L)-\Big[\FR{3}{2}C(C-A)+\FR{9}{4}H^2+m^2\Big]f(L)=0.
\ede
Changing variable from $L$ to $z=\cos^2(HL/2)$, and redefine $\wt{f}(z)=f(z)/\sqrt{1-z}$, we get an equation for $\wt{f}$,
\bge
  z(1-z)\wt{f}''(z)+(2-5z)\wt{f}'(z)-(4+m^2/H^2)\wt{f}(z)=0,
\ede
and the solution is
\bge
  \wt{f}(z)\propto{}_2F_1\big(2-\ii m/H,2+\ii m/H;2;z\big).
\ede
Therefore, after switching to original variable $f$ and imbedding coordinate $Z=\cos(HL)$, we have
\bge
  f(Z)\propto\sqrt{1-Z}{\,}_2F_1\Big(2-\FR{\ii m}{H},2+\FR{\ii m}{H};2;\FR{1+Z}{2}\Big).
\ede
To fix the prefactor, we calculate the corresponding correlators in flat space. In flat space,
\bge
  \la \Psi(x)\ob{\Psi}(y)\ra=\bgp \la\psi_\al\psi^\ga\ra & \la\psi_\al\psi_{\dot \ga}^\dag\ra \\ \la \psi^{\dag\dot\al}\psi^\ga\ra & \la \psi^{\dag \dot\al}\psi_{\dot\ga}^\dag \ra \edp=\int\FR{\di^4p}{(2\pi)^4}e^{\ii p\cdot(x-y)}(-p_\mu\ga^\mu+m).
\ede
We focus on the $\la\psi_\al\psi^\ga\ra=g(L)\de_\al{}^\ga$ element and carry out the integral, which gives
\bge
  g(L)=m\int\FR{\di^4p}{(2\pi)^4}e^{\ii p\cdot(x-y)}=\FR{m^2}{4\pi^2L}K_{1}(m L)\To \FR{m}{4\pi^2L^2}.
\ede
Then,
\bge
  f(L)=\FR{\ii}{m}g'(L)=-\FR{\ii}{2\pi^2L^3}.
\ede
On the other hand, $f(Z)$ approaches the following expression when $Z=\cos(HL)\to 1$,
\bge
  f(L)\To \FR{8\sqrt{2}}{\Gamma(2-\ii m/H)\Gamma(2+\ii m/H)(HL)^3}.
\ede
Compare the two, we get
\bge
  f(Z)=-\FR{\ii H^3\Gamma(2-\ii m/H)\Gamma(2+\ii m/H)}{16\sqrt{2}\pi^2}\sqrt{1-Z}{\,}_2F_1\Big(2-\FR{\ii m}{H},2+\FR{\ii m}{H};2;\FR{1+Z}{2}\Big),
\ede
and therefore,
\begin{align}
  g(Z)=&~\FR{\ii H}{m}\sqrt{1-Z^2}\bigg[\FR{\di}{\di Z}-\FR{3}{2(1-Z)}\bigg]f(Z)\n\\
  =&~\FR{H^3\Gamma(2-\ii m/H)\Gamma(2+\ii m/H)}{32\sqrt 2\pi^2}\FR{m}{H}\sqrt{1+Z}{\,}_2F_1\Big(2-\FR{\ii m}{H},2+\FR{\ii m}{H};3;\FR{1+Z}{2}\Big).
\end{align}

\section{Weyl Spinors on Slow-Roll Background}
\label{app_chp}
In this appendix, we supply several intermediate steps in deriving the mode functions (\ref{uplus})-(\ref{vminus}) in the presence of the chemical potential. Substituting the mode expansion (\ref{mode}) into the equation of motion (\ref{eqomcp}), we see that the modes satisfy the following equations
\begin{align}
  \ii\Big(\bar\si^{0\dot\al\be}\xi_{\be\pm}'+\ii\bar\si^{i\dot\al\be}k_i \xi_{\be\pm}\Big)=&~a\lam\bar\si^{0\dot\al\be}\xi_{\be\pm}+am\chi^{\dag\dot\al}_{\pm},\\
  \ii\Big(\bar\si^{0\dot\al\be}\chi_{\be\pm}'-\ii\bar\si^{i\dot\al\be}k_i \chi_{\be\pm}\Big)=&~a\lam\bar\si^{0\dot\al\be}\chi_{\be\pm}+am\xi^{\dag\dot\al}_{\pm}.
\end{align}
The second equation above can be rewritten as
\bge
  \ii\Big(-\chi^{\al}_\pm{}'\si^0_{\al\dot\be}+\ii k_i\chi^\al_\pm\si^i_{\al\dot\be}\Big)=-a\lam\chi^\al_\pm\si^0_{\al\dot\be}+am\xi_{\dot\be\pm}^\dag.
\ede
Taking the conjugate, we then get
\bge
  -\ii\Big(-\si^0_{\be\dot\al}\chi_\pm^{\dag\dot\al}{}'-\ii k_i\si_{\be\dot\al}^i\chi_\pm^{\dag\dot\al}\Big)=-a\lam\si^0_{\be\dot\al}\chi_\pm^{\dag\dot\al}+am\xi_{\be\pm}.
\ede
We further rewrite the mode functions in terms of helicity eigenspinors as in (\ref{helimode}), and get a pair of equations for $u$ and $v$,
\begin{align}
  &\ii u_\pm' \pm k u_\pm = a\lam u_\pm + am v_\pm ,\\
  &\ii v_\pm'\mp k v_\pm =-a\lam v_\pm +am u_\pm.
\end{align}
These two equations can be decoupled into a pair of second-order differential equations
\begin{align}
  &u_\pm''-a H u_\pm'+\Big[(\pm k-a\lam)^2+a^2m^2\pm\ii aH k\Big]u_\pm = 0,\\
  &v_\pm ''-a H v_\pm'+\Big[(\mp k+a\lam)^2+a^2m^2\mp\ii aH k\Big]v_\pm = 0,
\end{align}
which are readily solved in terms of Whittaker functions $W_{\ka,\mu}(z)$, as listed in (\ref{uplus})-(\ref{vminus}). The solution $W_{\ka,\mu}(z)$ is selected assuming the Bunch-Davis initial condition, i.e., $\psi\propto e^{-\ii k\tau}$ rather than $e^{+\ii k\tau}$ when $\tau\to-\infty$. The normalization, then, can be determined from the correctly normalized modes in the flat-space limit when $\tau\to-\infty$. A subtle point here is that the flat-space modes with no chemical potential do not contain a piece like $(-\tau)^{\pm\ii\lam}$ which appears in the early-time limit of mode functions. To avoid this issue, we can alternatively determine the normalization directly from the canonical commutation relation, $[\wt\psi_\al(\tau,\mb x),\wt\pi^\be(\tau,\mb y)]_+=\ii\de_\al{}^\be\de^{(3)}(\mb x-\mb y)$, where the conjugate momentum $\wt\pi^\be(\tau,\mb x)\equiv \de\ld/\de \wt\psi_\be'(\tau,\mb x) =\ii\wt\psi_{\dot \al}^\dag\bar\si^{0\dot\al\be}$. This implies the normalization of the mode functions,
\bge
  \sum_s\Big[\xi_{\al,s}\xi_{\dot \al,s}^\dag+\chi_{\dot \al,s}^\dag\chi_{\al,s}\Big]\bar\si^{0\dot\al\be}=\de_\al{}^\be.
\ede
In this calculation and others in the main text, the following several properties of the Whittaker function are useful. Firstly, we have a connection formula,
\bge
  W_{\ka,\mu}(z)=W_{\ka,-\mu}(z).
\ede
Then, the ``late-time'' limit,
\begin{align}
  &W_{\ka,\mu}(z)=\bigg[\FR{\Gamma(2\mu)}{\Gamma(\frac{1}{2}+\mu-\ka)}z^{1/2-\mu}+(\mu\to-\mu)\bigg]+\mathcal{O}(z^{3/2-\text{Re}\,\mu}),
  &&0\leq\,\text{Re}\,\mu<\FR{1}{2},~\mu\neq 0,
\end{align}
and the ``early-time'' limit,
\begin{align}
  &W_{\ka,\mu}(z)\sim e^{-z/2}z^\ka,
  &&|\text{ph}\;z|< \FR{3}{2}\pi
\end{align}
The derivatives of the Whittaker function
\begin{align}
  &zW_{\ka,\mu}'(z)=\Big[\Big(\FR{1}{2}-\ka\Big)^2-\mu^2\Big]W_{\ka-1,\mu}(z)-\Big(\FR{z}{2}-\ka\Big)W_{\ka,\mu}(z),\\
  &zW_{\ka,\mu}'(z)=-W_{\ka+1,\mu}(z)+\Big(\FR{z}{2}-\ka\Big)W_{\ka,\mu}(z).
\end{align}
Finally, the Wronskian
\bge
  \mathscr{W}\{ W_{\ka,\mu}(z),W_{-\ka,\mu}(e^{\pm\pi\ii}z)\}=W_{\ka,\mu}(z)\FR{\di}{\di z}W_{-\ka,\mu}(e^{\pm\pi\ii}z)-W_{-\ka,\mu}(e^{\pm\pi\ii}z)\FR{\di}{\di z}W_{\ka,\mu}(z)=e^{\mp\ka\pi\ii}.
\ede

\section{Schwinger-Keldysh Formalism with Weyl Spinors}
\label{app_sk}

Generally, the primordial non-Gaussianity as calculated in this paper is extracted from the $n$-point correlation functions $\la\de\phi\cdots\de\phi\ra$ of the inflaton perturbation $\de\phi$, evaluated at the late time limit $\tau=0$. The appropriate method for calculating these objects is the well-known in-in formalism, which can be viewed as a double copy of ``in-out formalism'', the standard procedure of calculating $S$-matrices. A classic approach to the in-in formalism is from the Hamiltonian formalism. See \cite{Chen:2010xka,Wang:2013eqj} for comprehensive reviews. On the other hand, the path-integral approach and the associated diagrammatic expansion, or Schwinger-Keldysh formalism, has been recently shown advantageous in several ways. We refer the readers to \cite{Chen:2017ryl} for a detailed review of path-integral based Schwinger-Keldysh formalism with applications in cosmological correlation functions. Here we spell out several new features of the formalism when applied to the two-component spinor fields.

As is well known, the path-integral variables for fermions are anticommutative Grassmannian numbers. When constructing the double-copy path integral in the in-in formalism, it is important to make sure that the anti-time-ordered contour is placed to the left of the time-ordered contour. Let the path-integral variables for the time-ordered and anti-time-ordered path integrals be $\psi_+$ and $\psi_-$, respectively, and let the corresponding external sources be $I_+$ and $I_-$, we can write down the generating functional of the in-in correlators as
\begin{align}
  Z[I_-,I_-^{\dag};I_+,I_+^\dag]=&~\int\Di\psi_-^\dag\Di\psi_-\exp\bigg[-\ii S[\psi_-,\psi_-^\dag]-\ii\int\di^4x\,\big(I_-^\al\psi_{-\al}^{}+I_{-\dot\al}^\dag\psi_-^{\dag\dot\al}\big)\bigg]\n\\
  &~\times\int\Di\psi_+^\dag\Di\psi_+\exp\bigg[\ii S[\psi_+,\psi_+^\dag]+\ii\int\di^4x\,\big(I_+^\al\psi_{+\al}^{}+I_{+\dot\al}^\dag\psi_+^{\dag\dot\al}\big)\bigg].
\end{align}
Here we have treated left-handed spinors $\psi_{\pm\al}$ and the right-handed spinors $\psi_\pm^{\dag\dot\al}$ as independent variables, and it is understood that the boundary conditions $\psi_+ = \psi_-$ and $\psi_+^\dag = \psi_-^\dag$ at the future infinity $\tau=0$ are imposed.

As an illustration of using the generating functional, here we derive all types of two-point correlation functions of the Weyl spinors $\psi$ and $\psi^\dag$, namely the propagators, by taking two functional derivatives with respect to corresponding external sources,
\begin{align}
  D_{ab\al\be}(k;\tau_1,\tau_2)=&~\int\di^3\mb X\,e^{-i\mb k\cdot\mb X}\FR{\de}{\ii a \de I_a^\al(\tau_1,\mb x)}\FR{\de}{\ii b \de I_b^\be(\tau_2,\mb y)}Z[I_-,I_-^{\dag};I_+,I_+^\dag]\bigg|_{I=0},\\
  D_{ab\al}{}^{\dot\be}(k;\tau_1,\tau_2)=&~\int\di^3\mb X\,e^{-i\mb k\cdot\mb X}\FR{\de}{\ii a \de I_a^\al(\tau_1,\mb x)}\FR{\de}{\ii b \de I_{b\dot\be}^\dag(\tau_2,\mb y)}Z[I_-,I_-^{\dag};I_+,I_+^\dag]\bigg|_{I=0}.
\end{align}
Then we can find all types of propagators as follows. Firstly, we have $\la\phi\phi^\dag\ra$-type propagators,
\begin{align}
  D_{++\al}{}^{\dot\be} (k;\tau_1,\tau_2)=&~\xi_\al(\tau_1,k)\xi^{\dag\dot\be}(\tau_2,k)\theta(\tau_1-\tau_2)-\chi^{\dag\dot\be}(\tau_2,k)\chi_\al(\tau_1,k)\theta(\tau_2-\tau_1),\\
  D_{+-\al}{}^{\dot\be}(k;\tau_1,\tau_2)=&-\chi^{\dag\dot\be}(\tau_2,k)\chi_\al(\tau_1,k),\\
  D_{-+\al}{}^{\dot\be}(k;\tau_1,\tau_2)=&~\xi_\al(\tau_1,k)\xi^{\dag\dot\be}(\tau_2,k),\\
  D_{--\al}{}^{\dot\be}(k;\tau_1,\tau_2)=&-\chi^{\dag\dot\be}(\tau_2,k)\chi_\al(\tau_1,k)\theta(\tau_1-\tau_2)+\xi_\al(\tau_1,k)\xi^{\dag\dot\be}(\tau_2,k)\theta(\tau_2-\tau_1).
\end{align}
In addition, we also have $\la\phi\phi\ra$-type propagators,
\begin{align}
  D_{++\al\be}(k;\tau_1,\tau_2)=&~\xi_\al(\tau_1,k)\chi_\be(\tau_2,k)\theta(\tau_1-\tau_2)-\xi_\be(\tau_2,k)\chi_\al(\tau_1,k)\theta(\tau_2-\tau_1),\\
  D_{+-\al\be}(k;\tau_1,\tau_2)=&-\xi_\be(\tau_2,k)\chi_\al(\tau_1,k),\\
  D_{-+\al\be}(k;\tau_1,\tau_2)=&~\xi_\al(\tau_1,k)\chi_\be(\tau_2,k),\\
  D_{--\al\be}(k;\tau_1,\tau_2)=&-\xi_\be(\tau_2,k)\chi_\al(\tau_1,k)\theta(\tau_1-\tau_2)+\xi_\al(\tau_1,k)\chi_\be(\tau_2,k)\theta(\tau_2-\tau_1).
\end{align}

\end{appendix}

\end{document}